\documentclass[aps,physrev,reprint,superscriptaddress,longbibliography]{revtex4-2}

\usepackage{comment}
\usepackage{siunitx}
\DeclareSIUnit{\rad}{rad}
\DeclareSIUnit{\langmuir}{L}
\DeclareSIUnit{\ML}{ML}
\DeclareSIUnit\angstrom{\text {Å}}
\usepackage{amsmath}
\usepackage{textcomp}
\usepackage{nicefrac} 
\usepackage{graphicx}
\usepackage{pifont}
\usepackage{placeins}
\usepackage{makecell}          
\usepackage{booktabs}
\usepackage{adjustbox} 
\usepackage{tikz}
\usepackage{pgfplots}
\pgfplotsset{compat=1.18}
\usepackage{multirow}

\usepackage[hidelinks, unicode=true, breaklinks=false, pdfborder={0 0 1}, backref=false, colorlinks=true, linkcolor=blue, urlcolor=blue, citecolor=blue]{hyperref}
\usepackage{microtype}
\begin{document}

\preprint{}

\title{Charge-to-spin conversion in epitaxial and polycrystalline Bi and Bi/Ag layers}

\author{Federica Nasr}
\email[]{federica.nasr@mat.ethz.ch}
\thanks{these authors contributed equally to this work.}
\affiliation{Department of Materials, ETH Zurich, CH-8093 Zurich, Switzerland}

\author{Emir Karadža}
\email[]{emir.karadza@mat.ethz.ch}
\thanks{these authors contributed equally to this work.}

\affiliation{Department of Materials, ETH Zurich, CH-8093 Zurich, Switzerland}

\author{Santos F. Alvarado}
\affiliation{Department of Materials, ETH Zurich, CH-8093 Zurich, Switzerland}

\author{Federico Binda}
\affiliation{Department of Materials, ETH Zurich, CH-8093 Zurich, Switzerland}

\author{Tobias Goldenberger}
\affiliation{Department of Materials, ETH Zurich, CH-8093 Zurich, Switzerland}

\author{Carlo Zucchetti}
\affiliation{Department of Materials, ETH Zurich, CH-8093 Zurich, Switzerland}
\affiliation{Dipartimento di Fisica, Politecnico di Milano, I-20133 Milano, Italy}

\author{Myriam H. Aguirre}
\affiliation{Instituto de Nanociencia y Materiales de Aragón, CSIC-Universidad de Zaragoza, ES-50009 Zaragoza, Spain}
\affiliation{Laboratorio de Microscopías Avanzadas, Universidad de Zaragoza, ES-50018 Zaragoza, Spain}
\affiliation{Dept. Física de la Materia Condensada, Universidad de Zaragoza, ES-50009 Zaragoza, Spain}

\author{Paolo Moras}
\affiliation{CNR-Istituto di Struttura della Materia (CNR-ISM), I-34149 Trieste, Italy}
\author{Andrey V. Matetskiy}
\affiliation{CNR-Istituto di Struttura della Materia (CNR-ISM), I-34149 Trieste, Italy}
\author{Polina M. Sheverdyaeva}
\affiliation{CNR-Istituto di Struttura della Materia (CNR-ISM), I-34149 Trieste, Italy}

\author{Paul Noël}
\email[]{paul.noel@ipcms.unistra.fr}
\affiliation{Department of Materials, ETH Zurich, CH-8093 Zurich, Switzerland}
\affiliation{Université de Strasbourg, CNRS, Institut de Physique et Chimie des Matériaux de Strasbourg, UMR 7504, F-67000 Strasbourg, France}

\author{Pietro Gambardella}
\email[]{pietro.gambardella@mat.ethz.ch}
\affiliation{Department of Materials, ETH Zurich, CH-8093 Zurich, Switzerland}

\date{\today}

\begin{abstract}
Bi is predicted to be an efficient generator of \mbox{spin--orbit} torques (SOTs), with charge-to-spin conversion efficiency comparable to those of prototypical heavy metals, such as Ta, W, and Pt. However, experimental reports provide widely scattered interconversion efficiencies, while the origin of the large conversion signal in Bi/Ag bilayers remains controversial. Here, we investigate charge-to-spin conversion in epitaxial and polycrystalline Bi-based magnetic heterostructures by measuring the damping-like SOT using magneto-optic Kerr effect magnetometry, complemented by structural and spectroscopic analyses and comparison with harmonic Hall resistance measurements. We show that inserting an Ag spacer between Bi(001) and metallic ferromagnets (FeCo or Ni) enhances the SOT efficiency by more than one order of magnitude, reaching an effective spin Hall conductivity of approximately 2\,$\times$\,$10^5\,(\hbar/2e)\,\SI{}{S/m}$, in excellent agreement with theoretical expectations for bulk Bi. This enhancement can be consistently explained by the preservation of the structural and chemical integrity of Bi\,$-$\,otherwise compromised by the direct deposition of a ferromagnetic overlayer\,$-$\,rather than by Rashba \mbox{spin--orbit} coupling at the Bi/Ag interface. We show that Ag forms an atomically sharp, chemically inert interface with Bi that, unlike other metallic spacers such as Al and Cu, prevents interdiffusion and solid-state dewetting while enabling efficient charge-to-spin conversion. Angle-resolved photoemission spectroscopy reveals no measurable enhancement of the Rashba spin splitting at the Bi(001) surface upon Ag deposition. Comparative studies across epitaxial, polycrystalline, and intentionally surface-oxidized Bi films, beyond oxygen doses known to destroy Bi(001) surface states, reveal that structural disorder has a negative impact on the SOT efficiency. Meanwhile, the retention of 72\% of the SOT efficiency following oxygen exposure indicates a dominant bulk contribution to spin-current generation in Bi/Ag heterostructures, yielding an effective Bi spin Hall angle of approximately 1. By establishing a direct correlation between atomic-scale integrity and charge-to-spin conversion, this study provides design principles to improve the reliability of Bi-based SOT devices and offers a robust framework for interpreting \mbox{spin--charge} interconversion in Bi and Bi/Ag systems.
\end{abstract}


\maketitle

\section{Introduction}
\vspace{-0.3cm}

Current-induced \mbox{spin--orbit} torques (SOTs) provide a versatile tool to develop magnetic technologies for information processing and storage~\cite{Manchon|RevModPhys|2019, Krizakova|JMMM|2022}. These technologies exploit materials capable of efficient charge-to-spin or charge-to-orbital conversion to manipulate the magnetization of magnetic layers and nanostructures. Such conversions arise from the spin Hall effect (SHE)~\cite{Sinova|RevModPhys|2015}, the spin \mbox{Rashba--Edelstein} effect (REE)~\cite{Bihlmayer|NatRevPhys|2022, Gambardella|PhilTransRSoc|2011}, and their orbital counterparts~\cite{Atencia|AdvinPhysX|2024,Jo|npjSpin|2024}. While the SHE stems from bulk \mbox{spin--orbit} coupling, the REE emerges from spin-momentum locking of the Fermi contour at the surface of topological materials and at the so-called Rashba interfaces, characterized by spin-split electronic states. This phenomenon, known as Rashba \mbox{spin--orbit} coupling, arises from the interplay between atomic \mbox{spin--orbit} interaction, which ties the electron’s spin to its orbital angular momentum, and the non-centrosymmetric crystal field created by broken inversion symmetry, which lifts the orbital degeneracy and generates a momentum- and band-dependent orbital angular momentum \cite{Bihlmayer|NatRevPhys|2022}.

Bi is the heaviest non-radioactive element and exhibits pronounced \mbox{spin--orbit} coupling effects in its electronic band structure~\cite{Hofmann|ProgSurfSci|2006,Guo|JMMM|2022}, making it a highly attractive material for realizing spin injectors and detectors. According to theoretical studies, the spin Hall conductivity (SHC) of semimetallic Bi is large and comparable to that of Pt, with predicted values spanning $(0.9$$-$$2.1)$$\times$$10^5(\hbar/2e)\,\SI{}{S/m}$, the latter corresponding to a spin Hall angle (SHA) of 0.24~\cite{Sahin|PhysRevLett|2015, Guo|JMMM|2022, Qu|PhysRevB|2023}. The SHC of Bi is also expected to be as large as that of the topological insulator material Bi$_\textrm{0.83}$Sb$_{0.17}$~\cite{Sahin|PhysRevLett|2015}. The emergence of Rashba interfaces in multilayer systems may further enhance the \mbox{spin--charge} interconversion efficiency beyond theoretical expectations for bulk Bi, as they introduce additional inversion-symmetry breaking. 

Despite a general consensus on the theoretical side that Bi should exhibit highly efficient \mbox{spin--charge} interconversion, experimental studies have reported widely scattered results. 
First, the measured SHA of Bi ranges from nearly negligible (0$-$0.02)~\cite{Hou|ApplPhysLett|2012,Emoto|JApplPhys|2014, Zhang|JApplPhys|2015, Sangiao|ApplPhysLett|2015, Nomura|ApplPhysLett|2015, Yue|PhysRevLett|2018, Emoto|PhysRevB|2016, Yue|APLMater|2021} to exceptionally large values, exceeding 0.5 and 5 in Refs.~\citenum{Kim|AdvSci|2023} and \citenum{Chi|PhysRevB|2022}, respectively. Second, while Bi-rich \mbox{Bi--Sb} alloys consistently show a large SHA, even when charge-to-spin conversion is promoted by thermally excited bulk carriers rather than topological surface states~\cite{Chi|SciAdv|2020, Binda|AdvMater|2023, Ou|NanoLett|2025}, this sizable SHA is not reliably reproduced in pure Bi~\cite{Sanchez|NatCommun|2013, Zhang|JApplPhys|2015, Sangiao|ApplPhysLett|2015, Nomura|ApplPhysLett|2015, Yue|PhysRevLett|2018, Emoto|PhysRevB|2016, Yue|APLMater|2021}, despite the similarity in the band structure between the two materials. Third, recent experimental studies of epitaxial Bi films reported a highly anisotropic SHA, varying from $-0.027$ in Bi(001) to 0.37 in Bi(012)~\cite{Liang|PhysRevB|2022, Fukumoto|PNAS|2023}. This is in stark contrast to the relatively small anisotropy reported in Ref.~\citenum{Kim|AdvSci|2023} as well as in theoretical studies~\cite{Guo|JMMM|2022, Qu|PhysRevB|2023}. 

Beyond its bulk-related spin Hall properties, Bi has also garnered considerable attention in the context of interface-driven \mbox{spin--orbit} coupling effects \cite{Marcano|PhysRevB|2010,Zucchetti|PhysRevB|2018}. In fact, the first signature of a large spin-to-charge conversion via inverse REE in a fully metallic system was reported in a Bi/Ag bilayer using spin pumping~\cite{Sanchez|NatCommun|2013}.
This discovery sparked extensive research into the Bi/Ag system via different techniques, e.g., spin pumping~\cite{Zhang|JApplPhys|2015, Sangiao|ApplPhysLett|2015, Nomura|ApplPhysLett|2015}, spin-polarized positron beams~\cite{Zhang|PhysRevLett|2015}, spin-torque ferromagnetic resonance~\cite{Jungfleisch|PhysRevB|2016}, magnetoresistance~\cite{Nakayama|PhysRevLett|2016}, and THz emission~\cite{Jungfleisch|PhysRevLett|2018, Zhou|PhysRevLett|2018}. However, conflicting findings emerged, including weak spin-to-charge conversion in the presence of ferromagnetic insulators~\cite{Yue|PhysRevLett|2018, Yue|APLMater|2021} and bulk-dominated rather than interface-driven transport~\cite{Matsushima|ApplPhysLett|2017, Shen|PhysRevLett|2021}.
Moreover, the small conversion efficiency reported in Bi single layers relative to Bi/Ag~\cite{Sanchez|NatCommun|2013, Zhang|JApplPhys|2015, Sangiao|ApplPhysLett|2015, Nomura|ApplPhysLett|2015} reinforces the idea of interface-driven conversion, but is not compatible with the large SHC expected in Bi.

In a recent study~\cite{Cheng|PhysRevLett|2022}, it was proposed that the inverse REE arises at the interface between Bi/Ag and metallic ferromagnets (FMs) and is strongly influenced by the ferromagnetic layer. However, the films used in this study suffered from strong interlayer mixing, as Bi was detected in both the capping Ag and the Ni overlayer. This behavior reflects the low sticking coefficient and the inherent diffusive tendency of Bi~\cite{Ueda|ThinSolidFilms|2020,Mahatha|SurfSci|2022} which stem from its low melting point ($\approx$\,$\SI{271}{\celsius}$) and relatively high volatility. As a result, Bi can segregate across both magnetic and nonmagnetic layers~\cite{Binda|PhysRevB|2021, Niimi|PhysRevLett|2012}, leading to the formation of interfacial~\cite{Vaughan|PhysRevRes|2020} and bulk~\cite{Cheng|PhysRevLett|2022} alloys. The extent of this intermixing is highly sensitive to parameters such as growth conditions, substrate choice, capping layers, and strain gradients~\cite{Kumari|ApplSurfSci|2007, Wang|IUCrJ|2020, Karbivskyy|ProgPhysMet|2021, Yaginuma|SurfSci|2007, Wansorra|JVacSciTechnolA|2021, Kumar|SurfEng|2021, Wansorra|ActaMater|2020}. Overall, these factors complicate the fabrication of well-defined Bi-based heterostructures and pose significant challenges to probing the intrinsic magnetotransport properties of Bi. Remarkably, alloying Bi with a small percentage of Sb significantly enhances its thermal and structural stability, reducing surface roughness and allowing the growth of more robust heterostructures~\cite{Walker|PhysRevMater|2019,Khaled|ACSApplElMater|2024,Ueda|ThinSolidFilms|2020}. Against this backdrop, addressing the correlation between structural and spin–charge interconversion properties of Bi-based multilayers becomes paramount, as this interplay has been largely overlooked.  

Here, we investigate charge-to-spin conversion in quasi-epitaxial (001)-oriented and polycrystalline Bi single layers and Bi/Ag bilayers interfaced with different metallic FMs using magneto-optic Kerr effect (MOKE) magnetometry, an accurate technique free of spurious magnetothermal and magnon-induced signals~\cite{Noel|PhysRevB|2025}. By combining scanning transmission electron microscopy (STEM), energy-dispersive X-ray spectroscopy (EDX), X-ray diffraction (XRD), angle-resolved photoemission spectroscopy (ARPES), and MOKE magnetometry, we establish a clear link between the structural, electronic, and magnetotransport properties of various Bi-based heterostructures. Our study demonstrates that both the sign and magnitude of the damping-like SOT efficiencies are critically governed by the materials adjoining the Bi layer, which strongly influences its structural integrity. In particular, we find that Bi films are highly susceptible to solid-state dewetting, alloying, oxidation, and crystallographic transformation when directly interfaced with metallic FMs such as FeCo and Ni. Similar effects occur upon contact with nonmagnetic metals such as Cu and Al. Conversely, interfacing Bi with an Ag overlayer suppresses these structural degradation pathways, preserving the smoothness and continuity of the Bi layer. As a result, the effective SHC in Bi/Ag/FM is enhanced by more than one order of magnitude compared to Bi/FM, reaching values of ${(1.9\pm0.1)}$ and ${(0.70\pm0.03)}$$\times$$10^5(\hbar/2e)\,\SI{}{S/m}$ for quasi-epitaxial and polycrystalline Bi layers, respectively, consistent with theoretical predictions for bulk Bi~\cite{Guo|JMMM|2022}. The sign of the SHC in Bi/Ag/FM is positive, similar to Pt. Conversely, the SHC has a negative sign in Bi/FeCo, which we attribute to negative self-induced magnetic torques in the ferromagnetic layer. In Bi(001)/Ni, pronounced \mbox{Bi--Ni} alloying leads to a reduced (yet positive) effective SHC compared to Bi(001)/Ag/Ni. Inverting the stacking order, that is, depositing Bi on top of Ni, directly stabilizes a (012) texture and suppresses interlayer mixing, yielding a positive SHC in Ni/Bi that remains largely unaffected by the deposition of an Ag overlayer. 
Altogether, these observations indicate that Ag serves as a spin-transparent diffusion barrier that prevents catastrophic modifications of the Bi layer when in direct contact with other metals, in particular ferromagnetic metals. The large charge-to-spin conversion efficiency obtained with an Ag interlayer is largely associated with a preserved sample integrity, rather than an enhanced \mbox{Rashba--Edelstein} contribution in the Bi/Ag system. Our claim is further supported by ARPES measurements, which reveal that Ag deposition does not affect the Rashba spin splitting at the surface states of Bi(001). Further exposure of Bi(001) to an oxygen dose sufficient to destroy its surface states~\cite{Liu|PhysE|2012} prior to Ag deposition reduces the SOT efficiency by only about $28\%$, providing compelling evidence for bulk-dominated spin generation. Therefore, our results show that the previously reported discrepancies on \mbox{spin--charge} interconversion in Bi films could be solely caused by a strong variability in the structural quality of the samples.

This paper is organized as follows. Section~\ref{sec:methods} introduces the experimental techniques employed in this work. Section~\ref{sec:growth} details the epitaxial growth, structural characterization, and stabilization of Bi(001) thin films on BaF$_2$(111) (see also Note 1 of the Supplemental Material~\cite{SI}). Section~\ref{sec:ARPES} discusses the ARPES spectrum of Bi(001) and the influence of Ag adatoms. Section~\ref{sec:structural_magnetic}, complemented by Supplementary Note 2 of the Supplemental Material~\cite{SI}, provides a comprehensive structural and magnetic characterization of Bi(001)/Ag/FM heterostructures with and without Ag spacer. Section~\ref{sec:SOT} reports the MOKE measurements of the damping-like SOT in the different samples, with further details provided in Note 3 of the Supplemental Material~\cite{SI}, and includes a comparison with relevant literature. The estimated torque efficiencies are compared to those obtained from harmonic Hall resistance (HHR) measurements in Note 4 of the Supplemental Material~\cite{SI}. Section~\ref{sec:SOT_alternative} extends the study to (012)-textured and polycrystalline Bi-based heterostructures, while Sec.~\ref{subsec:spacers} and Note 5 of the Supplemental Material~\cite{SI} address Bi(001)/X/FeCo, with X\,=\,Al and Cu. The results are discussed and further correlated with literature reports in Sec.~\ref{sec:discussion}, followed by the conclusions in Sec.~\ref{sec:conclusion}. A consolidated summary of the structural, magnetic, and charge-to-spin conversion properties for all investigated samples is provided in Fig.~\ref{fig:summary} below, facilitating direct comparison and correlation across the different heterostructures.

\section{Methods}
\label{sec:methods}
\vspace{-0.3cm}

\subsection{Thin-film deposition}
\label{sec:methods_growth}
\vspace{-0.3cm}

The films were deposited by molecular beam epitaxy (MBE) in an ultra-high vacuum (UHV) system with a base pressure of ${10^{-10}\,\SI{}{mbar}}$. BaF$_2(111)$ and MgO(001) substrates from SurfaceNet GmbH were cleaned by sequential sonication in (i) acetone and (ii) isopropanol (for BaF$_2$) or ethanol (for MgO), with each step lasting $\SI{5}{min}$. After cleaning, the substrates were degassed in UHV at \SI{500}{\celsius} for $\SI{60}{min}$. High-purity elements ($\geq99.999\%$) were evaporated while maintaining the chamber pressure in the $10^{-10}\,$mbar range. The growth rates were calibrated prior to deposition using quartz crystal microbalances and kept approximately constant across samples: Bi on MgO and BaF$_2$ (\SI{3.0}{\angstrom/min}), Bi on Ni (\SI{1.4}{\angstrom/min}), Ag (\SI{1.3}{\angstrom/min}), Cu (\SI{1.0}{\angstrom/min}), Fe (\SI{0.6}{\angstrom/min}), Co (\SI{0.6}{\angstrom/min}), Ni (\SI{1.2}{\angstrom/min}), and Al (\SI{1.2}{\angstrom/min}). For Bi growth on top of BaF$_2$ and MgO (Ni), the substrates were precooled to $\SI{-30}{\celsius}$ ($\SI{-190}{\celsius}$) and the evaporation was initiated within $\SI{5}{min}$ of transfer into a separated chamber. Bi was deposited at oblique incidence of the molecular beam, following the procedure described in Ref.~\citenum{Binda|AdvMater|2023}, which is known to suppress crystal twinning in Bi$_{0.9}$Sb$_{0.1}$(001) films. However, unlike Bi$_{0.9}$Sb$_{0.1}$(001), full suppression of twinning was not achieved in Bi(001) \{see Fig.~S1(a) within the Supplemental Material~\cite{SI}\}. After growth, the Bi films were annealed at approximately $\SI{200}{\celsius}$ for \SI{45}{min}. Ag, Cu, and Al were deposited on Bi at $\SI{-190}{\celsius}$ to suppress interlayer diffusion and minimize surface roughness. Fe and Co were co-deposited at room temperature from separate rod-fed electron-beam sources. Ni films on MgO were deposited at room temperature and subsequently annealed at \SI{350}{\celsius} for \SI{45}{min}. All layers other than Bi were deposited with the molecular beam oriented at normal incidence. In multilayers, no post-deposition annealing was performed to avoid intermixing and alloying, unless otherwise specified. To prevent sample oxidation upon air exposure, a protective aluminum oxide (AlO$_x$) capping layer was deposited in two steps. First, an initial $\SI{10}{\angstrom}$ of Al was grown in UHV at room temperature. Each sample was then exposed to molecular oxygen at $\SI{1.6e-6}{mbar}$ for \SI{5}{min}, after which an additional $\SI{25}{\angstrom}$ of Al was deposited under the same oxygen pressure. Post-growth sample characterization via X-ray techniques was performed using a Panalytical X'Pert MRD diffractometer for X-ray reflectivity (XRR) and azimuthal angle scans and a Panalytical X'Pert PRO MPD diffractometer for XRD ${\theta/2\theta}$ scans, employing Cu K-$\alpha$1 radiation.

\vspace{-0.3cm}
\subsection{Device fabrication}
\label{sec:methods_fabrication}
\vspace{-0.3cm}

Hall bar devices with a width of $\SI{10}{\micro\meter}$ and a length of $\SI{100}{\micro\meter}$ were fabricated using optical lithography and Ar ion milling. Each device features three Hall branches, equally spaced by \SI{50}{\micro\meter}. In the first step, UV photolithography was used to define photoresist patches matching the geometry of the Hall bars. The unprotected regions were then etched down to the substrate using Ar ion milling. Residual photoresist was removed through repeated oxygen plasma ashing cycles. A second photolithography step was performed to pattern photoresist trenches for the electrical contacts. Metallic contact pads, consisting of Cr[\SI{10}{nm}]/Au[\SI{200}{nm}], were deposited by thermal evaporation, followed by lift-off in acetone to complete the device fabrication. 

\vspace{-0.3cm}
\subsection{Scanning transmission electron microscopy}
\label{sec:methods_TEM}
\vspace{-0.3cm}

High-resolution high-angle annular dark-field STEM (HAADF-STEM) imaging was performed using a FEI Titan G2 60-300 transmission electron microscope operated at \SI{300}{kV}, equipped with a CEOS probe aberration corrector, and offering subnanometer resolution of \SI{0.8}{nm}.
Cross-sectional TEM lamellae were prepared using a FEI Helios 650 DualBeam FIB system with a Ga ion source. The focused ion beam was initially operated at \SI{30}{kV} and progressively reduced below \SI{5}{kV} in the final milling steps to minimize surface damage. 

\vspace{-0.3cm}
\subsection{Angle-resolved photoemission spectroscopy}
\label{sec:methods_ARPES}
\vspace{-0.3cm}

ARPES measurements were carried out at the VUV-Photoemission beamline of the Elettra synchrotron in Trieste, Italy, using horizontally polarized light with a photon energy of \SI{50}{\electronvolt}. The sample was kept at $\SI{17}{\kelvin}$ during the measurements. The experimental end-station was equipped with a Scienta R-4000 electron spectrometer located at a $45^\circ$ angle relative to the incident photon beam. The samples were prepared \emph{in situ} using an integrated MBE system with a base pressure of $2$\,$\times$\,$10^{-10}\,\SI{}{mbar}$, equipped with low-energy electron diffraction apparatus for surface characterization.

\vspace{-0.3cm}
\subsection{Magneto-optic Kerr effect}
\label{sec:methods_MOKE}
\vspace{-0.3cm}

We used polar MOKE to measure the SOT in nonmagnet/ferromagnet (NM/FM) heterostructures~\cite{Fan|NatCommun|2014, Montazeri|NatCommun|2015, Noel|PhysRevB|2025}. This technique relies on measuring the SOT-induced oscillations of the magnetization upon injection of an alternating current of amplitude $I$ into a Hall-bar-shaped sample. For a current directed along $\textbf{x}$, a nonequilibrium spin accumulation with polarization $\boldsymbol{\sigma}$\,$\parallel$\,$\textbf{y}$ is generated at the NM/FM interface via mechanisms such as the SHE and REE. Absorption of the spin accumulation by the FM results in a damping-like SOT~\cite{Manchon|RevModPhys|2019}, which can be represented via an effective field $\textbf{B}_\textrm{DL}$\,=\,$B_\textrm{DL} \,\boldsymbol{\sigma}$\,$\times$\,$\textbf{m}$, where $\textbf{m}$ is the unit magnetization vector~\cite{Garello|NatNano|2013}. In the linear-response regime, if the magnetization is aligned parallel or antiparallel to the current direction by an external magnetic field of amplitude $H_{\rm x}$, $\textbf{B}_\textrm{DL}$ drives small out-of-plane oscillations of the magnetization along $\textbf{z}$, at the modulation frequency of the injected current. In devices with homogeneous magnetic, interfacial, and current-flow properties, this effective field is spatially uniform across the current channel~\cite{Fan|NatCommun|2014, Noel|PhysRevB|2025}. Additionally, the Oersted field induces out-of-plane magnetization tilts of opposite sign at opposite edges of the Hall bar. For thin-film geometries ($t \ll w$, with $t$ the thickness and $w$ the width of the current line), the $z$ component of the Oersted field can be analytically estimated using the Biot-Savart law~\cite{Silva|JApplPhys|1999,Hayashi|PhysRevB|2014,Fan|NatCommun|2014},
\begin{equation}\label{eq:BOe_MOKE}
    H_\textrm{Oe}^z(y) = -\dfrac{I} {2\pi\,w}\,\ln\left[{\dfrac{w-\left(\frac{w}{2}+y\right)}{\frac{w}{2}+y}}\right].
\end{equation}
Unlike the damping-like effective field, the Oersted field does not depend on the orientation of the magnetization and is solely defined by the charge current direction. Therefore, the two fields can be distinguished by their opposite symmetry with respect to magnetization reversal.

In this work, we used a linearly polarized laser beam with a wavelength of \SI{520}{nm} focused at normal incidence on the sample surface and a spot diameter of approximately $\SI{0.5}{\micro\meter}$ at full width half maximum. The samples were mounted on nanopositioners that allowed for high-precision scanning of the laser spot over the device surface. The electrical current was modulated at a frequency of $\SI{8751}{Hz}$. The polar MOKE geometry makes the experiment exclusively sensitive to the out-of-plane components of the magnetization. The rotation of the light polarization upon reflection from the sample because of changes in the magnetization direction, namely the Kerr rotation $\theta_K$, was measured using a balanced photodetector and a lock-in amplifier~\cite{Stamm|PhysRevLett|2017}. The magnetization oscillations are detected as the first harmonic component of $\theta_K$, making the MOKE technique largely insensitive to magnetothermal effects, which instead contribute to the second harmonic response.

The Kerr rotation components arising from the Oersted and damping-like effective fields are given by~\cite{Fan|NatCommun|2014}:
\begin{subequations} \label{eq:MOKE}
    \begin{equation}
        \theta_{K}^\textrm{Oe} = \dfrac{\theta_K(H_\mathrm{x})+\theta_K(-H_\mathrm{x})}{2},
        \label{eq:Kerr_Oe}
    \end{equation}
    \begin{equation}
        \theta_{K}^\textrm{DL} = \dfrac{\theta_K(H_\mathrm{x})-\theta_K(-H_\mathrm{x})}{2}.   
        \label{eq:Kerr_DL}
    \end{equation}
\end{subequations}
By scanning the laser beam along the $y$ axis, the spatial profile of $\theta_K$ was captured.  Measurements were performed while successively reversing the external magnetic field direction between $\pm{\textbf{x}}$. The quantities $\theta_K^\textrm{Oe}$ and $\theta_K^\textrm{DL}$ were obtained by symmetrizing the data according to Eq.~\eqref{eq:MOKE}. By fitting the functional dependence given by Eq.~\eqref{eq:BOe_MOKE} to the experimental $\theta_K^\text{Oe}$ profile, the Kerr rotation (in units of $\si{\micro\rad}$) was directly related to the corresponding effective field (in units of mT), yielding the calibration constant (magneto-optical constant) of the measured system. As both the damping-like and the Oersted fields cause a tilt of the magnetization in the out-of-plane direction, the extracted magneto-optical constant applies to both contributions. Consequently, $B_\textrm{DL}$ was obtained by fitting a constant line to the spatial profile of $\theta_{K}^\textrm{DL}$, excluding data near the device edges to avoid spurious boundary effects. The negligible contribution of a quadratic MOKE signal~\cite{Montazeri|NatCommun|2015} was verified via field-scan measurements \{see Fig.~S4(d) within the Supplemental Material~\cite{SI}\}. A detailed description of the MOKE setup and torque evaluation procedure used in this study can be found in Refs.~\citenum{Noel|PhysRevB|2025} and \citenum{Stamm|PhysRevLett|2017}.

\section{Results}
\vspace{-0.3cm}
\subsection{Growth of quasi-epitaxial Bi(001)}
\label{sec:growth}
\vspace{-0.3cm}

\begin{figure*}[t!]
\centerline{\includegraphics[width=1.2\textwidth]{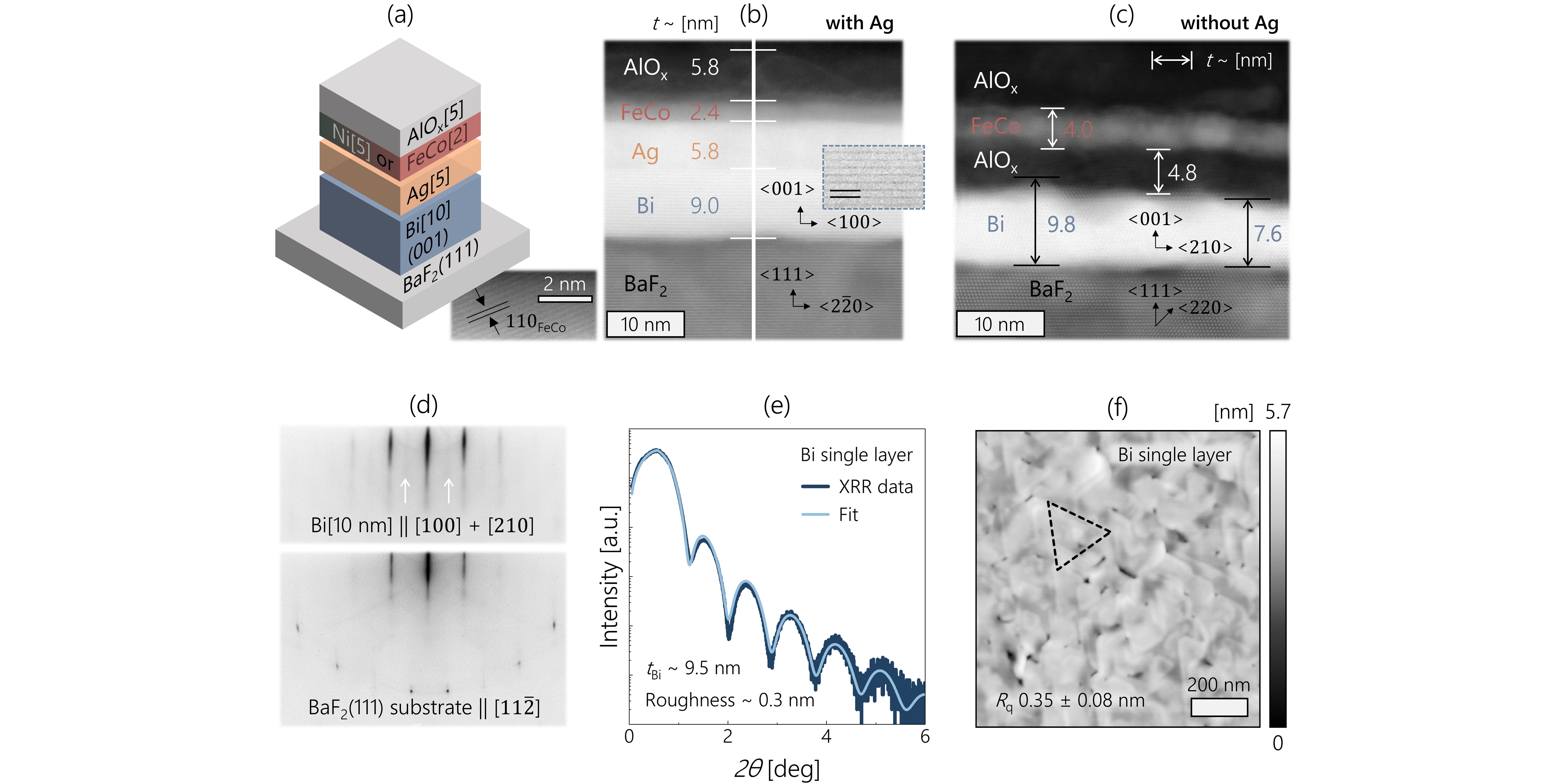}}
\caption{\label{fig:figure1} Structural characterization of Bi(001)-based heterostructures grown on BaF$_2$(111). (a) Schematic of the Bi(001)/Ag/FM stack. (b, c) HAADF-STEM images of (b) Bi[10]/Ag[5]/FeCo[2] and (c) Bi[10]/FeCo[2]. The lamellae were cut from the Hall bar devices used to perform SOT measurements. (d) RHEED pattern of a 10-nm-thick Bi film after annealing (top) and of the BaF$_2$(111) substrate prior to Bi deposition (bottom). The data were acquired with an electron beam energy of 30 keV. The white arrows are guides to the eye for identifying the residual Bi(012) reflexes. (e) XRR of a 10-nm-thick uncapped Bi film. (f) AFM Z-sensor image of a 10-nm-thick uncapped Bi film. All thicknesses are expressed in [nm].}
\end{figure*}  

We fabricated ferromagnetic heterostructures of the type BaF$_2$(111)/Bi(001)[10]/Ag[5]/FM[$t_\textrm{FM}$]/AlO$_x$, hereinafter referred to as Bi(001)/Ag/FM, with and without an Ag spacer, using MBE (see details in Sec.~\ref{sec:methods_growth}). Here, the thicknesses are expressed in [nm], whereas the crystallographic orientation of Bi is expressed according to the three-index hexagonal notation. The ferromagnetic layer consists of either Fe$_{0.5}$Co$_{0.5}$[2] or Ni[5]. 
Figures~\ref{fig:figure1}(a) and~\ref{fig:figure1}(b) show a schematic of the Bi/Ag/FM stack and a cross-sectional HAADF-STEM image of the representative sample Bi(001)/Ag/FeCo, respectively. We note that to account for possible sample alteration during device fabrication, the STEM samples were prepared from Hall bar devices rather than from the full films (see details in Sec.~\ref{sec:methods_TEM}). The HAADF-STEM image reveals a sharp interface between Bi and Ag, with no major interdiffusion between the two layers. In addition, the thicknesses of the individual layers are in good agreement with the nominal values provided in Fig.~\ref{fig:figure1}(a). From electron diffraction analysis, we deduce that all layers are quasi-epitaxial [see insets of Fig.~\ref{fig:figure1}(b), highlighting the atomic planes of Bi and FeCo], with the [001] axis of Bi oriented out of the plane and a calculated in-plane lattice constant of the hexagonal unit cell $a_\textrm{Bi}$\,$\approx$\,$\SI{4.54}{\angstrom}$, which is close to the bulk value~\cite{Hofmann|ProgSurfSci|2006}. HAADF-STEM results are in excellent agreement with the XRD ${\theta/2\theta}$ scans performed on the full film, as reported in Fig.~\ref{fig:figure3}(b). In particular, reflexes from the crystal planes (111) and (222) of BaF$_2$, (003), (006), and (009) of Bi, and (220) of Ag confirm the epitaxial relationship between the substrate and the different layers. In this regard, the absence of any reflexes from FeCo can be ascribed to the small thickness of the film and the high background signal arising from the X-ray fluorescence of Fe-containing materials irradiated with Cu radiation~\cite{Mos|GeomicrobiolJ|2018}. From the position of the (006) Bi peak ($2\theta$\,=\,$45.69^{\circ}$) and the angular spacing between the Laue oscillations, we estimated an out-of-plane lattice constant of the hexagonal unit cell $c_\textrm{Bi}$\,$\approx$\,$\SI{11.9}{\angstrom}$ and an out-of-plane crystallite size of approximately $\SI{8.85}{nm}$, respectively \{see Fig.~S1(b) within the Supplemental Material~\cite{SI}\}.
Interestingly, the HAADF-STEM image of the nominal Bi(001)/FeCo sample in Fig.~\ref{fig:figure1}(c) reveals that removing the Ag spacer leads to a pronounced degradation of the film structure. In particular, HAADF-STEM analysis shows regions in which the film continuity is completely disrupted \{see Fig.~S2(a) within the Supplemental Material~\cite{SI}\}, as well as continuous areas exhibiting a pronounced Bi surface roughness (exceeding \SI{2}{nm}) together with the formation of an additional interlayer between Bi and FeCo. As a result, the nominal Bi(001)/FeCo stacking sequence is no longer preserved. Complementary EDX analysis identifies this additional interlayer as AlO$_x$ and also indicates substantial Bi oxidation. The corresponding EDX maps for the Bi(001)/Ag/FeCo and Bi(001)/FeCo samples are shown in Fig.~S2 within the Supplemental Material~\cite{SI}.

To gain more insight into the structural properties of Bi single layer grown on BaF$_2$(111), we performed \textit{in situ} RHEED and \textit{ex situ} atomic force microscopy (AFM) and XRR on uncapped Bi single layers. Figure~\ref{fig:figure1}(d) presents the RHEED pattern of Bi (top) and BaF$_2$(111) prior to Bi deposition (bottom) for the illustrative case of the electron beam oriented along the [$11\bar2$] crystallographic axis of the substrate. The RHEED pattern acquired on the postannealed Bi film exhibits a sixfold azimuthal symmetry, consistent with the (001) crystallographic orientation of the film. From the horizontal spacing between the reflexes, we estimate a lattice mismatch between Bi(001) and BaF$_2$(111) of approximately $3.2 \%$, which confirms that the Bi film is not compressively strained but relaxed to its bulk lattice constant \{see Fig.~S1(c) within the Supplemental Material~\cite{SI} for the supporting reciprocal space mapping data\}. The presence of faint extra-reflexes (highlighted by white arrows) is indicative of residual (012)-oriented crystal grains in the film, not fully suppressed by the post-growth thermal treatment, confirming the quasi-epitaxial growth of Bi. The elongation of the reflexes in the RHEED pattern provides a first indication of the smoothness of the Bi film. The roughness and root-mean-square roughness extracted from XRR and AFM measurements in Figs.~\ref{fig:figure1}(e) and~\ref{fig:figure1}(f) are $R_q$\,=\,$0.35\pm\SI{0.08}{\nm}$ and $R_{rms}$\,$\approx$\,$\SI{0.3}{\nm}$, respectively. These values are lower than the roughness values of 2, 3, and \SI{3.5}{\nm} reported in the literature for 8-nm-thick Bi films thermally evaporated on YIG(111)~\cite{Yue|PhysRevLett|2018}, Al$_2$O$_3$(0001)~\cite{Zhang|PhysRevLett|2015}, and SiO$_2$~\cite{Sanchez|NatCommun|2013}, respectively. Also, the film is smoother than the 15-nm-thick sputtered Bi in Ref.~\citenum{Kim|AdvSci|2023}, with estimated roughness of \SI{3}{\nm} from AFM and \SI{5}{\nm} from XRR. A comparable roughness of \SI{0.4}{\nm} was reported for a 31.6-nm-thick Bi film sputter-deposited on a YIG slab at low temperature~\cite{Yue|APLMater|2021}.

\vspace{-0.3cm}
\subsection{Electronic structure of Bi(001)/Ag}
\label{sec:ARPES}
\vspace{-0.3cm}

The BiAg$_2$ substitutional surface alloy, formed by depositing $\nicefrac{1}{3}$ monolayer (ML) of Bi on Ag(111), exhibits the largest Rashba spin splitting reported to date~\cite{Ast|PhysRevLett|2007}. However, the reciprocal configuration\,$-$\,namely, the deposition of Ag on Bi\,$-$\,remains largely unexplored. The BiAg$_2$ alloy tends to form on the surface of Ag layers grown on top of a BiAg$_2$-terminated Ag(111) crystal, owing to the surfactant behavior of Bi~\cite{Mahatha|SurfSci|2022}. A similar tendency could manifest when Ag is deposited on top of a Bi(001) crystal, thereby substantiating the microscopic origin of the giant spin-to-charge conversion efficiency in the Bi/Ag bilayer~\cite{Sanchez|NatCommun|2013}.

To directly test this scenario, we performed ARPES on Bi(001) films coated with varying sub-monolayer to monolayer thicknesses of Ag ($\SI{1}{ML}$\,=\,$\SI{2.36}{\angstrom}$). The photon energy of \SI{50}{eV} employed in these experiments provides the highest surface sensitivity, since electrons photoemitted near the Fermi energy (the region of interest in this case) have kinetic energies of approximately \SI{45}{eV}, where the inelastic mean free path curve for electrons in solids has its absolute minimum ($\lambda$\,=\,$\SI{0.5}{nm}$)~\cite{Seah|SufInterfaceAnal|1979}. This means that approximately 99\% of the photoelectrons that have not undergone inelastic scattering events within the material are generated within $\SI{1.5}{nm}$ of the Bi film surface. This condition allows us to capture in detail the changes induced by Ag deposition on the electronic structure of the Bi film.

\begin{figure*}[t!]
\centerline{\includegraphics[width=1.2\textwidth]{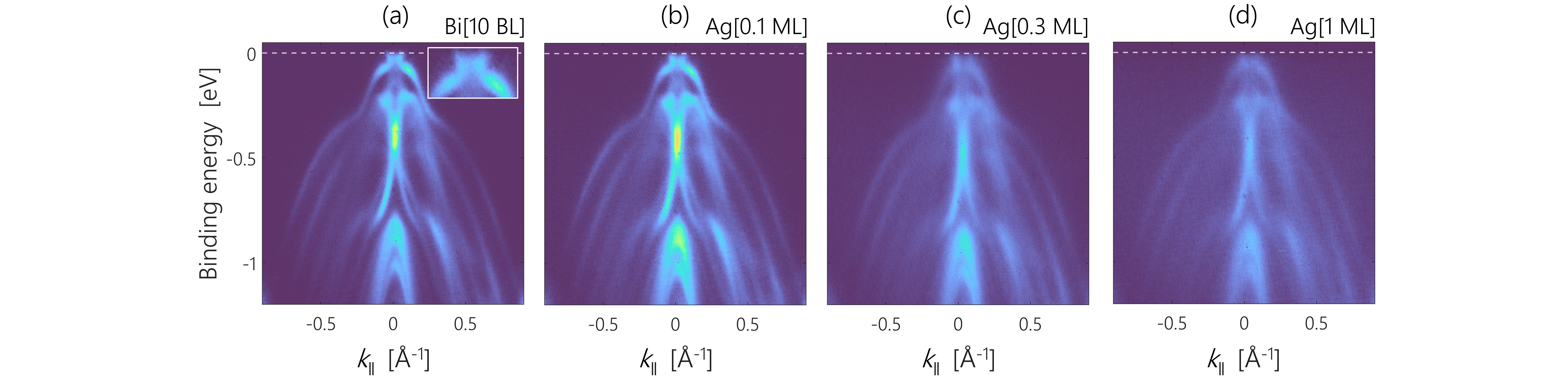}}
\caption{\label{fig:figure2} ARPES spectra. 
(a-d) Electronic dispersion of (a) 10-BL-thick Bi(001) grown on $7\times7-$Si(111) before Ag deposition and after deposition of (b) \SI{0.1}{\ML}, (c) \SI{0.3}{\ML}, and (d) \SI{1}{\ML} of Ag (1~BL of Bi\,=\,\SI{3.93}{\angstrom}, \SI{1}{ML} of Ag\,=\,\SI{2.36}{\angstrom}). The inset in (a) shows a zoom on the Rashba-split surface state of Bi(001) at the Fermi level. The spectra were acquired at a temperature of \SI{17}{\kelvin} and a photon energy of \SI{50}{\electronvolt} along the ${\bar{\Gamma}-\bar{\textrm{K}}}$ direction of the first Brillouin zone of Bi(001). The data were treated with a Fourier filter to remove the grid structure from the experimental setup.}
\end{figure*}

Figure~\ref{fig:figure2}(a) shows the band dispersion of a pristine 4-nm-thick Bi(001) film grown on a $(7\times7)$-reconstructed Si(111) substrate, along the $\bar{\Gamma}-\bar{\textrm{K}}$ direction of the surface Brillouin zone. The inset highlights the Bi spin-polarized surface states near the Fermi level. The measured band structure is consistent with previous reports~\cite{Hirahara|PhysRevB|2007, Bian|PhysRevB|2009}. Furthermore, in Fig.~\ref{fig:figure2}(a), quantum well states can be recognized following their characterization in Ref.~\citenum{Bian|PhysRevB|2009}, thus confirming the atomically defined nature of the surface and interface planes. Ag was deposited at liquid-nitrogen temperature to minimize surface roughness and suppress interdiffusion, followed by self-annealing at room temperature. Figures~\ref{fig:figure2}(b-d) show that, as the Ag coverage increases from 0.1 to \SI{1}{ML}, the dispersion of the Bi electronic states remains essentially unaltered, and no new electronic states emerge. The major change consists of an increased broadening of the photoemission peaks, indicating that Ag acts as a scattering source for the photoelectrons emitted from the Bi(001) film. Overall, this analysis demonstrates that the BiAg$_2$ surface alloy does not form upon Ag deposition on Bi(001).

\vspace{-0.3cm}
\subsection{Structural and magnetic properties of Bi(001)/Ag/FM and Bi(001)/FM heterostructures}
\label{sec:structural_magnetic}
\vspace{-0.3cm}

\begin{figure*}[t!]
\centerline{\includegraphics[width=1.2\textwidth]{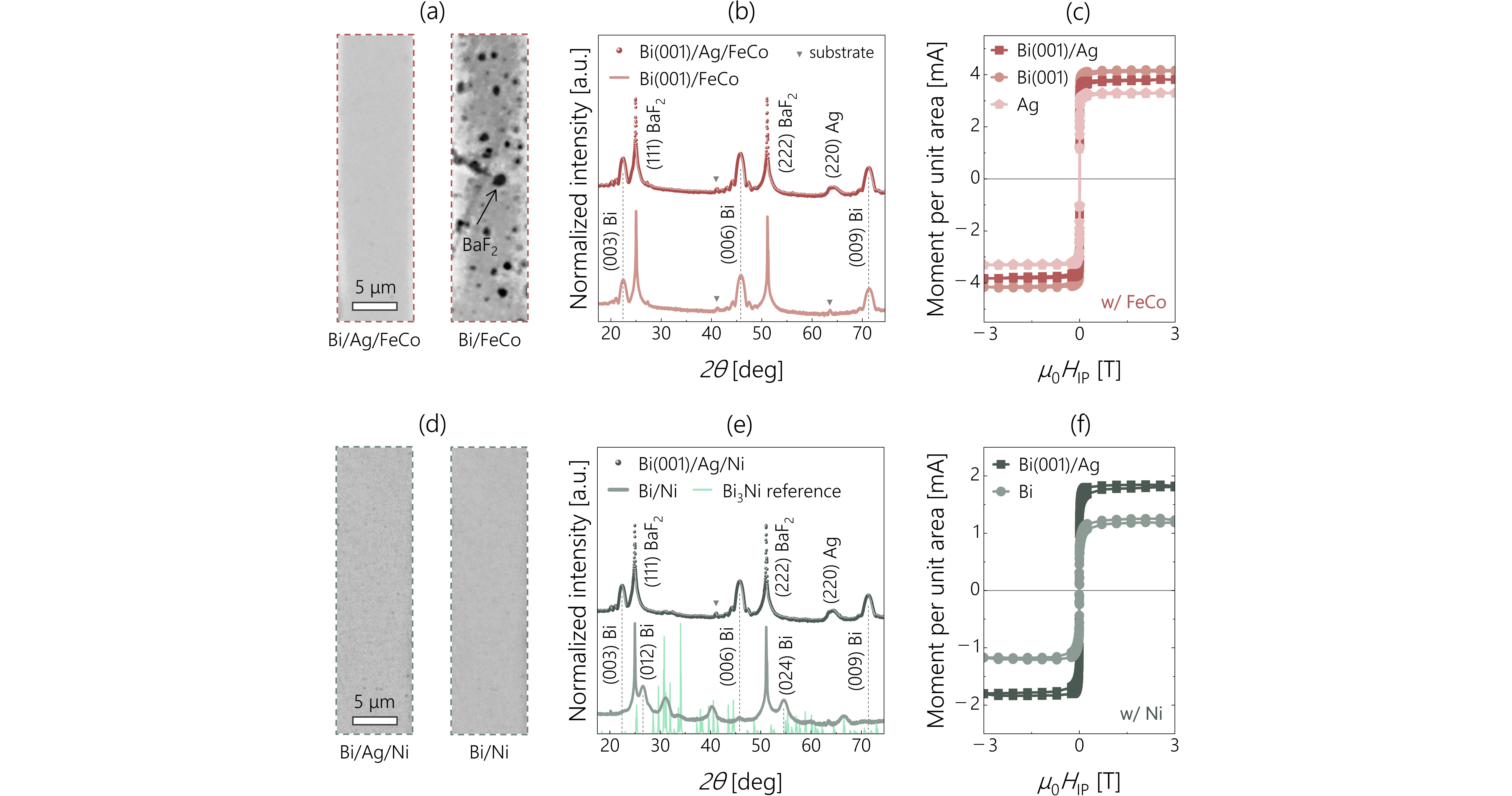}}
\caption{\label{fig:figure3} Structural and magnetic characterization of Bi(001)-based heterostructures grown on BaF$_2$(111).
(a) Optical microscope images of a section of the Hall bar device along the current line of Bi[10]/Ag[5]/FeCo[2] (left) and Bi[10]/FeCo[2] (right). (b) XRD ${\theta/2\theta}$ scans of Bi[10]/Ag[5]/FeCo[2] and Bi[10]/FeCo[2]. (c) Room-temperature magnetization curves measured by SQUID magnetometry for Bi[10]/Ag[5]/FeCo[2], Bi[10]/FeCo[2], and Ag[5]/FeCo[2]. (d) Optical microscope images of a section of the Hall bar device along the current line of Bi[10]/Ag[5]/Ni[5] (left) and Bi[10]/Ni[5] (right). (e) XRD ${\theta/2\theta}$ scans of Bi[10]/Ag[5]/Ni[5] and Bi[10]/Ni[5]. In the latter, the \mbox{Bi--Ni} interdiffusion causes the conversion of Bi(001)/Ni into Bi(012)/Bi$_3$Ni/Ni. The diffractogram of the Bi$_3$Ni alloy, shown as a reference, was retrieved from the Powder Diffraction File 00-054-0537 provided by the International Centre for Diffraction Data. (f) Room-temperature magnetization curves measured by SQUID magnetometry for Bi[10]/Ag[5]/Ni[5] and Bi[10]/Ni[5]. $\mu_0 H_\text{IP}$ in (c) and (f) denotes the external magnetic field applied in the sample plane. All thicknesses are expressed in [nm].}
\end{figure*}

We assessed the structural continuity of the multilayer stacks grown on the $\textrm{BaF}_{2}(111)$ substrate using optical microscopy, as shown in Figs.~\ref{fig:figure3}(a) and~\ref{fig:figure3}(d) for samples incorporating FeCo and Ni as ferromagnetic layers, respectively. Bi(001)/Ag/FeCo is continuous, consistent with the HAADF-STEM observations discussed in Sec.~\ref{sec:growth}. In contrast, direct deposition of FeCo onto Bi(001) disrupts the structural integrity of the Bi film, as evidenced by the visible pinholes or voids in the optical micrograph. The measured depth using AFM ranges from 10 to \SI{20}{nm}, depending on the probed surface region. These features are indicative of Bi solid-state dewetting, likely driven by interfacial strain developed during the growth of the FeCo overlayer (see Fig.~S3 within the Supplemental Material~\cite{SI} for supporting scanning electron microscopy measurements).
This finding is fully consistent with HAADF-STEM imaging of Bi(001)/FeCo, discussed in Sec.~\ref{sec:growth}, which shows film discontinuity, Bi roughening and oxidation, and interlayer mixing. Despite such pronounced morphological differences, the XRD ${\theta/2\theta}$ scans of Bi(001)/Ag/FeCo and Bi(001)/FeCo shown in Fig.~\ref{fig:figure3}(b) appear qualitatively similar. Minor distinctions are the lower diffraction peak intensity and the slightly faster attenuation of the Laue fringes in the Bi(001)/FeCo sample, indicative of reduced structural coherence in the absence of an Ag spacer. 

Interestingly, although Bi/Ni retains a continuous morphology, as shown in Fig.~\ref{fig:figure3}(d), XRD ${\theta/2\theta}$ measurements reported in Fig.~\ref{fig:figure3}(e) reveal substantial chemical interaction between Bi and Ni. Direct deposition of Ni onto Bi(001) leads to interlayer diffusion and the formation of a polycrystalline \mbox{Bi--Ni} alloy. This is evidenced by the appearance of multiple diffraction peaks in the XRD pattern of Bi/Ni, which can be attributed to the Bi$_3$Ni alloy based on their angular position. Additionally, the residual unalloyed Bi(001) converts into Bi(012). Effectively, the nominal Bi(001)/Ni stack transforms into a Bi(012)/Bi$_3$Ni/Ni structure, here referred to as Bi/Ni. Despite alloying, the overall film remains continuous, in contrast to the solid-state dewetting observed in Bi(001)/FeCo. Once again, these results underscore the stabilizing role of the Ag spacer, which preserves the structural and chemical integrity of Bi.

Further, we characterized the magnetic properties of the heterostructures using a superconducting quantum interference device (SQUID) at room temperature. Figures~\ref{fig:figure3}(c) and~\ref{fig:figure3}(f) show the magnetic moment per unit area as a function of the in-plane external magnetic field measured for FeCo- and Ni-based samples, respectively. Using the nominal thickness of the ferromagnetic layer, we can extract saturation magnetization values of \mbox{$M_s$\,=\,$\SI{1.87e6}{A/m}$} for Bi(001)/Ag/FeCo, \SI{2.05e6}{A/m} for Bi(001)/FeCo, and \SI{1.63e6}{A/m} for Ag/FeCo. These values are in good agreement with the bulk saturation magnetization reported for bcc Fe$_{0.5}$Co$_{0.5}$ of $\SI{1.95e6}{A/m}$~\cite{Coey|MMM|2010}. The smaller $M_s$ of Ag/FeCo can be attributed to common effects observed in ultra-thin ferromagnetic films, such as interfacial mixing with nonmagnetic elements and the formation of magnetically dead layers. Equivalently, we can extract \mbox{$M_s$\,=\,$\SI{3.54e5}{A/m}$} for Bi(001)/Ag/Ni, which is close to the bulk value reported for Ni of ${\SI{4.88e5}{A/m}}$~\cite{Coey|MMM|2010}. In contrast, Bi/Ni exhibits an approximately 34\% reduction in the measured magnetic moment per unit area (\SI{1.16}{mA}) with respect to Bi(001)/Ag/Ni (\SI{1.77}{mA}). This observation is consistent with the structural evidence of \mbox{Bi--Ni} interfacial mixing. Assuming that this reduction arises primarily from alloying-induced loss of magnetic volume, we can estimate an effective Ni thickness ${t_\textrm{Ni} \leq \SI{3.3}{nm}}$ in the Bi/Ni sample. To further corroborate this finding, we performed anomalous Hall measurements on patterned Hall bar devices and estimated an anomalous Hall resistance of \mbox{$R_\textrm{AHE}$\,=\,$\SI{0.07}{\ohm}$} for Bi(001)/Ag/Ni and $\SI{0.56}{\ohm}$ for Bi/Ni \{see Fig.~S7(a) within the Supplemental Material~\cite{SI}\}. Using a parallel resistor model, we further estimated the partial current flowing through the Bi and Ni layers in Bi(001)/Ag/Ni, which corresponds to about 45\% of the total current. In the ideal scenario of constant Ni thickness across the samples, we would expect the anomalous Hall resistance of Bi/Ni to scale as \mbox{$R_\textrm{AHE}^\textrm{Bi/Ni}$\,=\,$R_\textrm{AHE}^\textrm{Bi/Ag/Ni}/\,0.45$\,$\approx$\,$\SI{0.16}{\ohm}$}, which is about 3.6 times smaller than the experimentally measured value. Assuming the standard scaling relation \mbox{$R_\textrm{AHE}$\,=\,$\rho_\textrm{AHE}/t_\textrm{FM}$}, where $\rho_\textrm{AHE}$ is the anomalous Hall resistivity and $t_\textrm{FM}$ is the thickness of the ferromagnetic layer, we infer a residual Ni thickness of only $\SI{1.4}{nm}$ in Bi/Ni.

\vspace{-0.3cm}
\subsection{SOTs in Bi(001)/Ag/FM and Bi(001)/FM heterostructures}
\label{sec:SOT}
\vspace{-0.3cm}

To quantify the damping-like SOT in our samples, we employed scanning MOKE magnetometry (see Sec.~\ref{sec:methods_MOKE} and Refs.~\citenum{Fan|NatCommun|2014} and~\citenum{Montazeri|NatCommun|2015}). Unlike the HHR method~\cite{Avci|PhysRevB|2014}, MOKE detection is inherently insensitive to current-induced magnetothermal effects and magnonic contributions, providing a significant advantage for accurate SOT quantification~\cite{Noel|PhysRevLett|2025, Noel|PhysRevB|2025}. This is especially relevant in Bi-based heterostructures, where the combination of Bi's high carrier mobility and low carrier density promote pronounced Nernst effects~\cite{Behnia|ReprProgPhys|2016, Yue|PhysRevLett|2018}. The resulting transverse thermoelectric voltages exhibit the same symmetry with respect to the in-plane magnetization angle as the damping-like transverse voltage (see details in Note 4 of the Supplemental Material~\cite{SI}). Consequently, methods such as the HHR, which rely on performing in-plane scans of the magnetization direction, can fail to disentangle the overlapping signals, especially when the transverse voltage due to the ordinary Nernst effect and damping-like SOT are of comparable magnitude. Moreover, Ni is particularly prone to magnonic excitations because of its relatively low saturation magnetization and low Curie temperature among the 3$d$ ferromagnetic transition metals, further complicating the HHR analysis. A comprehensive comparison between MOKE and HHR is provided in Note 4 of the Supplemental Material~\cite{SI}. The results clearly reveal a substantial misestimation of the damping-like SOT magnitude in both FeCo- and Ni-based samples when the latter approach is used, underscoring the importance of employing a detection method that is robust against thermal and magnonic artifacts.

To measure current-induced SOTs, we patterned the sample in the form of Hall bar devices (see details in Sec.~\ref{sec:methods_fabrication}), as illustrated in Fig.~\ref{fig:figure4}(a). Here, the current injection direction defines the $x$-axis of the coordinate system, whereas $R^\textrm{L}$ represents the four-probe longitudinal resistance. A schematic cross-sectional view of the device is shown in the same Fig.~\ref{fig:figure4}(a), depicting a transverse cut along the current line and outlining the MOKE-based SOT detection method (see details in Sec.~\ref{sec:methods_MOKE}). Figures~\ref{fig:figure4}(b-e) show the optical reflectance as well as the antisymmetric ($\theta_K^\textrm{Oe}$) and symmetric ($\theta_K^\textrm{DL}$) components of the Kerr rotation for the various FeCo- and Ni-based samples, corresponding to the Oersted and damping-like contributions, respectively. For the Bi(001)/FeCo sample, the scans were carried out over a continuous section of the current line with no visible pinholes under the optical microscope, as evidenced by the nearly constant optical reflectance in Fig.~\ref{fig:figure4}(c). Fits to $\theta_K^\textrm{Oe}$ using Eq.~\eqref{eq:Kerr_Oe} and to $\theta_K^\textrm{DL}$ using a constant are shown as solid lines. Data shown in panel (b) are acquired with \SI{4}{mA} of total charge current flowing in the device, in (c) with \SI{2.5}{mA}, while in (d) and (e) with \SI{5}{mA}. Throughout this work, a positive $\theta_K^\textrm{DL}$ indicates a positive damping-like effective field, consistent with the convention used for Pt/Co bilayers (bottom/top configuration). Although the polarity of the Oersted field remains the same across the samples, we observe a clear sign reversal of the damping-like Kerr rotation for Bi(001)/FeCo. Specifically, $\theta_K^\textrm{DL}$ is positive in Bi(001)/Ag/FeCo and Bi(001)/Ag/Ni, negative in Bi(001)/FeCo, and again positive in Bi/Ni. This implies that removing the Ag spacer reverses the direction of the damping-like SOT in FeCo-based samples, but not in Ni-based ones. It is also important to note that, despite the Kerr rotation signals of Bi(001)/Ag/Ni and Bi/Ni in Figs.~\ref{fig:figure4}(d) and~\ref{fig:figure4}(e) were recorded under the same total current, Bi/Ni presents a marked decrease in both $\theta_K^\textrm{Oe}$ and signal-to-noise ratio compared to Bi(001)/Ag/Ni. In this regard, although $B^z_\textrm{Oe}$ as defined in Eq.~\eqref{eq:BOe_MOKE} depends only on $I$ and $w$, $\theta_K^\textrm{Oe}$ is also dependent on both the thickness and the magneto-optical constant of the ferromagnetic layer, where the former defines the magneto-optical interaction volume. Assuming similar magneto-optical constants for the two Ni-based samples, the degraded signal in Bi/Ni is a clear indication of inhomogeneity and reduced magnetic volume, further evidencing substantial \mbox{Bi--Ni} intermixing \{see also Fig.~S4(a) within the Supplemental Material~\cite{SI} for overlaid data\}. 
To have a more quantitative estimate of the damping-like effective field in Bi/Ni and account for possible inhomogeneity in the \mbox{Bi--Ni} intermixing across the Hall bar device, we further integrated the Kerr rotation signal recorded at different $x$ positions along the current line. With this procedure, a positive offset superimposed on the characteristic antisymmetric Kerr rotation profile expected for a homogeneous Oersted field could be obtained, indicating the presence of a damping-like SOT contribution with positive sign in Bi/Ni. 

\vspace{-0.3cm}
\subsubsection{SOT efficiencies and comparison with literature}
\vspace{-0.3cm}

\begin{figure*}[t!]
\centerline{\includegraphics[width=1.2\textwidth]{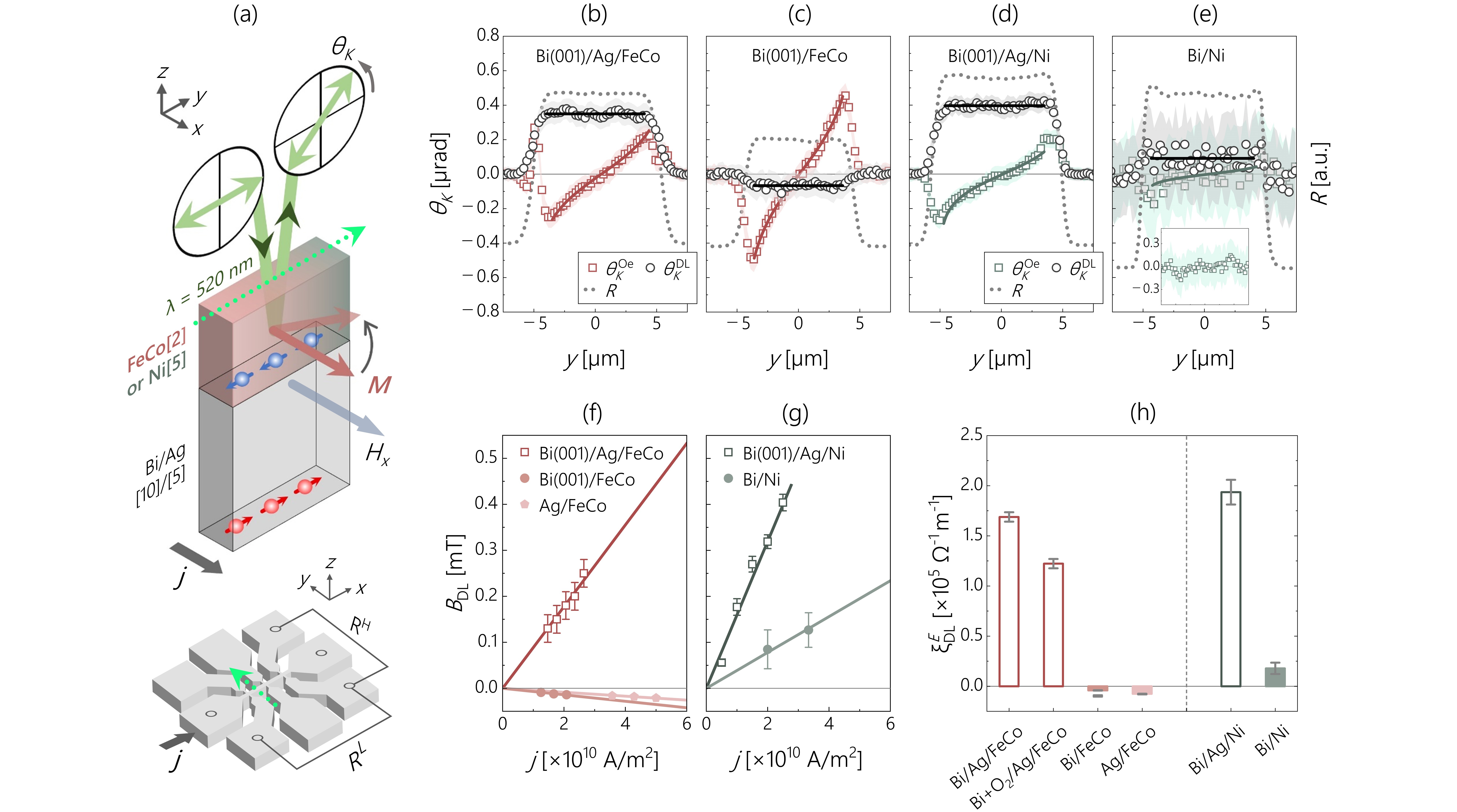}}
\caption{\label{fig:figure4} SOT measurements of Bi(001)-based heterostructures grown on BaF$_2$(111).
(a) Illustration of a Hall bar device with superimposed electrical connections and reference system (bottom) and schematic of the scanning MOKE magnetometry in polar geometry used to perform optical measurements of SOTs (top). A laser beam (green line) is focused onto the Hall bar device and scanned along a transverse cross section of the current line (indicated by the green dotted arrow). The Kerr rotation ($\theta_K$) caused by the out-of-plane magnetization oscillation induced by the Oersted field and the damping-like effective field is recorded. (b-e) MOKE detection of damping-like SOT in (b) Bi[10]/Ag[5]/FeCo[2] at $j$\,=\,$\SI{2.35e10}{A/m^2}$ ($I$\,=\,$\SI{4}{mA}$), (c) Bi[10]/FeCo[2] at \SI{2.08e10}{A/m^2} (\SI{2.5}{mA}) (d) Bi[10]/Ag[5]/Ni[5] at \SI{2.50e10}{A/m^2} (\SI{5}{mA}), and (e) Bi[10]/Ni[5] at \SI{3.33e10}{A/m^2} (\SI{5}{mA}). The data were acquired with an in-plane magnetic field $(\mu_0H_x)$ of \SI{80}{mT}. The gray dots represent the optical reflectance ($R$) of the sample, whereas the red/green squares and the black circles represent the Kerr rotation induced by the Oersted field ($\theta_K^\textrm{Oe}$) and the damping-like field ($\theta_K^\textrm{DL}$), calculated using Eqs.~\eqref{eq:Kerr_Oe} and \eqref{eq:Kerr_DL}, respectively. The continuous lines are fits to the data. (f, g) Damping-like effective field ($B_\textrm{DL}$) versus average current density ($j$) for the samples with FM equal to (f) FeCo and (g) Ni. The solid lines are linear fits to the data with intercept forced to zero. (h) Damping-like SOT efficiency ($\xi_\textrm{DL}^E$) with respect to the electric field ($E$) estimated for the different samples using the linear fits in (f) and (g). The errors are calculated from the uncertainty in the fits in (f) and (g) and in the saturation magnetic moment per unit area (${M_s\,t_\textrm{FM}}$). For Bi[10]/FeCo[2], results from two nominally identical samples are included. Positive SOT efficiencies indicate a positive SHC sign, consistent with that of Pt. All thicknesses are expressed in [nm].}
\end{figure*}

Figures~\ref{fig:figure4}(f) and~\ref{fig:figure4}(g) show the estimated damping-like effective field as a function of the total current density for FeCo- and Ni-based samples, respectively. 
To compare the strength of the damping-like SOT across different samples, we quantified the damping-like SOT efficiency with respect to the electric field $E$ as
\begin{equation}
    \xi^E_\textrm{DL} = \dfrac{2e}{\hbar}\dfrac{M_s\,t_\textrm{FM}\,B_\textrm{DL}}{E},
    \label{eq:xi_DL^E}
\end{equation}
expressed in units of [S/m], where ${M_st_\textrm{FM}}$ is the saturation magnetization per unit area extracted from SQUID magnetometry (see details in Sec.~\ref{sec:structural_magnetic}), whereas ${B_\textrm{DL}/E}$ is obtained from the slope of linear fits to the data in Figs.~\ref{fig:figure4}(f) and~\ref{fig:figure4}(g), with intercepts constrained to zero. $\xi^E_\textrm{DL}$ is commonly referred to as the effective SHC, where the term \textit{effective} acknowledges the influence of spin reflection and spin memory loss at the interface
between the nonmagnetic and the magnetic layer. In this sense, $\xi^E_\textrm{DL}$ represents a lower bound on the intrinsic charge-to-spin conversion capability of the nonmagnetic layer. Multiplication of $\xi^E_\textrm{DL}$ by $(\hbar/2e)$ yields the SHC in units of angular momentum, as often reported in the theoretical literature. Throughout this work, positive $\xi^E_\textrm{DL}$ values denote the same SHC sign as Pt.

We choose to normalize the damping-like effective field by the electric field rather than the current density in the nonmagnetic layer in order to avoid assumptions about current partitioning in the multilayer stack. This is particularly relevant given the stack-dependent microstructure, hence resistivity, of the Bi layer as well as the presence of interfacial conduction channels that are not captured by simplified parallel resistor models. Estimating the effective SHA as $\xi^{j_\textrm{NM}}_\textrm{DL}$\,=\,$\xi^E_\textrm{DL}/\sigma_\textrm{0}$
would require prior knowledge of both the microscopic origin of the SOT and the electrical conductivity $\sigma_0$ of the spin-converting layer. This conductivity may include contributions from bulk, interface, or alloyed regions. Such parameters are generally unknown or poorly defined in structurally complex systems like Bi-based heterostructures. Conversely, $\xi^E_\textrm{DL}$ provides a robust and assumption-free metric to compare SOT in systems with distinct structural and compositional properties. The estimated $\xi_\textrm{DL}^E$ values for the different samples are reported in Fig.~\ref{fig:figure4}(h) and further compiled in the summary figure (see Fig.~\ref{fig:summary}).

We find large and positive SOT efficiencies of $\xi^E_\textrm{DL}$\,=\,$(1.69 \pm 0.05)$\,$\times$\,$10^{5}\,\text{S/m}$ for Bi(001)/Ag/FeCo and $(1.9 \pm 0.1)$\,$\times$\,$10^{5}\,\text{S/m}$ for Bi(001)/Ag/Ni. These values are comparable to those of prototypical heavy metal systems used for the generation of SOT, such as Pt, with $\xi^E_\textrm{DL}$ in the range of $(2.43$$-$$3.5)$\,$\times$\,$10^{5}\,\text{S/m}$~\cite{Manchon|RevModPhys|2019}. Our values are also consistent with those reported for a Bi$_{0.9}$Sb$_{0.1}$(001)/FeCo bilayer of $(2.2\pm0.2)$\,$\times$\,$10^{5}\,\text{S/m}$~\cite{Binda|AdvMater|2023} and for Bi(001)/NiFe and Bi$_{0.84}$Sb$_{0.16}$(001)/NiFe bilayers of approximately 1.2\,$\times$\,$10^{5}\,\text{S/m}$~\cite{Ou|NanoLett|2025}. The SOT efficiencies of our Bi(001)/Ag/FM heterostructures significantly exceed those reported for sputter-deposited Bi/CoFeB bilayers of about 0.54\,$\times$\,$10^{5}\,\text{S/m}$~\cite{Kim|AdvSci|2023}, and align well with theoretical predictions for the intrinsic SHC of Bi~\cite{Sahin|PhysRevLett|2015, Guo|JMMM|2022}. By replacing the quasi-epitaxial Bi(001) with a polycrystalline Bi film, we observe a significant reduction of the SOT efficiency, which approaches the value provided in Ref.~\citenum{Kim|AdvSci|2023}. The results on polycrystalline Bi are further discussed in Sec.~\ref{sec:SOT_alternative}. Unlike Ref.~\citenum{Cheng|PhysRevLett|2022}, which reported more than an order of magnitude enhancement of the spin-to-charge conversion efficiency in YIG/Bi/Ag/FM upon replacing Fe with Ni, we do not find a strong dependence of the damping-like SOT on the ferromagnetic layer. In our study, replacing FeCo with Ni yields a modest increase of less than 15\% in ${\xi^E_\textrm{DL}}$.
This discrepancy likely originates from structural differences between the samples. Reference~\citenum{Cheng|PhysRevLett|2022} reports extensive Bi diffusion into both the Ag and Ni layers, which may drastically modify both bulk and interfacial \mbox{spin--orbit} coupling effects. In contrast, our HAADF-STEM imaging (see details in Sec.~\ref{sec:growth}) shows a sharp and well-defined Bi(001)/Ag interface, suggesting limited interlayer diffusion and preservation of the intrinsic electronic structure of Bi. 
Upon removal of the Ag spacer, the SOT efficiency decreases markedly to $\xi^E_\textrm{DL}$\,=\,$(0.18 \pm 0.06)$\,$\times$\,$10^{5}\,\text{S/m}$ in Bi/Ni and $(-0.068 \pm 0.003)$\,$\times$\,$10^{5}\,\text{S/m}$ in Bi(001)/FeCo, with the latter additionally exhibiting a sign reversal of the SOT. Because of the large signal-to-noise ratio affecting the MOKE data of Bi/Ni, the SOT efficiency was estimated from the integrated Kerr rotation signal, rather than from a single line scan, as discussed above. 
To investigate the origin of the negative SOT in Bi(001)/FeCo, we measured a reference sample consisting of an Ag/FeCo bilayer grown on BaF$_2$(111), which yields a similarly negative value of $(-0.078 \pm 0.001)$\,$\times$\,$10^{5}\,\text{S/m}$. The agreement in both magnitude and sign suggests that the damping-like SOT in Bi(001)/FeCo and Ag/FeCo originates from self-torques within the FeCo layer, rather than from spin current generation in either Bi or Ag. This interpretation is consistent with previous studies reporting a negligible SHA for Ag~\cite{Sanchez|NatCommun|2013, Niimi|PhysRevB|2014}. It is interesting to note that the SOT efficiency of our Bi(001)/FeCo closely matches the value provided for an amorphous Bi$_{0.9}$Sb$_{0.1}$/FeCo bilayer of $(-0.060 \pm 0.002)$\,$\times$\,$10^{5}\,\text{S/m}$ in Ref.~\citenum{Binda|AdvMater|2023}. This study revealed a nearly two orders of magnitude reduction in $\xi^E_\textrm{DL}$ when transitioning from epitaxial to amorphous Bi$_{0.9}$Sb$_{0.1}$, accompanied by a sign reversal. Such a trend mirrors the evolution of the SOT efficiency in our Bi(001)-based samples upon removal of the Ag spacer, which leads to degradation of the Bi film. 

\vspace{-0.55cm}
\subsubsection{Influence of Bi surface oxidation} 
\label{subsec:oxidation}
\vspace{-0.3cm}
To assess the role of the Bi/Ag interfacial quality in generating and transmitting a spin current, we intentionally degraded the Bi(001) surface prior to Ag deposition by exposing it to an oxygen partial pressure of $\SI{1.6E-6}{mbar}$ for \SI{5}{min}, corresponding to a total oxygen dose of 360 Langmuirs (L). This experiment is inspired by ARPES measurements~\cite{Liu|PhysE|2012}, which show that the surface state band of Bi(001) vanishes beyond an O$_2$ exposure of $\SI{4.5}{\langmuir}$. The oxygen dose of $\SI{360}{\langmuir}$ was therefore deliberately chosen to exceed the threshold required to destroy the surface states of Bi, while preventing bulk oxidation or degradation. We verified the retention of the bulk structural properties of Bi using both \textit{in situ} and \textit{ex situ} structural characterization techniques. Immediately after oxygen exposure, the RHEED streaks of Bi(001) remained visible, albeit noticeably blurred, indicating that the film was not fully oxidized or amorphized within the near-surface region probed by the electron beam. Additionally, HAADF-STEM and EDX analyses revealed no measurable change in the structural quality of Bi and no evidence of bulk oxidation compared to the pristine sample.

Following oxygen exposure, we find that the SOT efficiency of Bi(001)+O$_2$/Ag/FeCo decreases by approximately 28\% compared to Bi(001)/Ag/FeCo. The substantial residual torque signal, measured despite the expected suppression of surface states, implies that the damping-like SOT in Bi(001)/Ag/FM heterostructures is not exclusively interface driven. Instead, it involves a dominant bulk-mediated contribution, such as spin current generation via the SHE in Bi and its subsequent diffusion through the spin-transparent Ag layer. We emphasize that the oxygen exposure does not enable a quantitative separation of surface state and bulk contributions, as surface oxidation may also reduce the interface spin transparency for spin Hall currents generated in the bulk or introduce additional spin memory loss.

\vspace{-0.3cm}
\subsubsection{Cu and Al spacers} 
\label{subsec:spacers}
\vspace{-0.3cm}
Notably, we find that replacing Ag with alternative metallic spacers, such as Cu or Al, does not preserve the structural integrity of the Bi(001) film. In both cases, similar to the scenario without any spacer, the Bi layer exhibits significant discontinuity \{see Fig.~S8(b) within the Supplemental Material~\cite{SI}\}. In Bi(001)/Cu/FeCo, EDX analysis reveals strong overlap between the Bi and Cu signals, suggesting \mbox{Bi--Cu} intermixing and the formation of a Bi-rich alloy \{see Fig.~S8(e) within the Supplemental Material~\cite{SI}\}. This overlap can be rationalized by considering that the electron beam crosses different sections of the lamella characterized by distinct local compositions, as result of solid-state dewetting and island formation: Bi/Cu/FeCo regions, where the film remains continuous, and Cu/FeCo regions that experienced local Bi film breakup during Cu deposition. The resulting EDX signal would therefore represent an average of these two local compositions, mimicking the appearance of \mbox{Bi--Cu} intermixing. This interpretation is consistent with the persistence of both the diffraction peaks of Bi in the XRD ${\theta/2\theta}$ scan \{see Fig.~S8(c) within the Supplemental Material~\cite{SI}\} and the Bi atomic planes in the HAADF-STEM image \{see Fig.~S8(d) within the Supplemental Material~\cite{SI}\}. The Bi(001)/Cu/FeCo sample exhibits an SOT efficiency of \mbox{$\xi^E_\textrm{DL}$\,=\,$(0.37 \pm 0.01)$\,$\times$\,$10^5\,\SI{}{S/m}$}, which is significantly reduced compared to Bi(001)/Ag/FeCo. Interestingly, the positive $\xi^E_\textrm{DL}$ contrasts with previous studies~\cite{Niimi|PhysRevLett|2012, Niimi|PhysRevB|2014, Tatsuoka|PhysRevB|2024} that report a negative SHA in Cu doped with Bi impurities. The observed discrepancy may arise from several possible mechanisms: (i) the Bi-rich \mbox{Bi--Cu} alloy, different from Bi-doped Cu samples, may intrisically show a positive SHA; (ii) the positive SOT contribution from residual, unalloyed Bi regions may partially offset the negative SOT contribution from \mbox{Bi--Cu} alloy, resulting in an overall positive effective SHC; (iii) a Bi concentration gradient at the interface between the \mbox{Bi--Cu} alloy and the Cu layer may give rise to interfacial \mbox{spin--orbit} coupling contributing to the measured SOT signal; and (iv) solid-state dewetting may promote edge or side oxidation, where oxidized Cu has previously been shown to exhibit significant \mbox{spin--charge} interconversion efficiency~\cite{An|NatCommun|2016}. Employing an Al spacer leads to even greater destabilization of the Bi film, resulting in a negligibly small SOT signal \{see Fig.~S8(f) within the Supplemental Material~\cite{SI}\}. 
We note that, similarly to Ag, light metals such as Cu and Al present very long spin diffusion lengths exceeding $\SI{100}{nm}$~\cite{Idzuchi|PhysE|2015}, ruling out reduced SOT efficiency coming from spin absorption in the metallic spacer.

\vspace{-0.3cm}
\subsection{Structural characterization and SOTs in polycrystalline and Bi(012)-based heterostructures}
\label{sec:SOT_alternative}
\vspace{-0.3cm}

\begin{figure*}[t!]
\centerline{\includegraphics[width=1.2\textwidth]{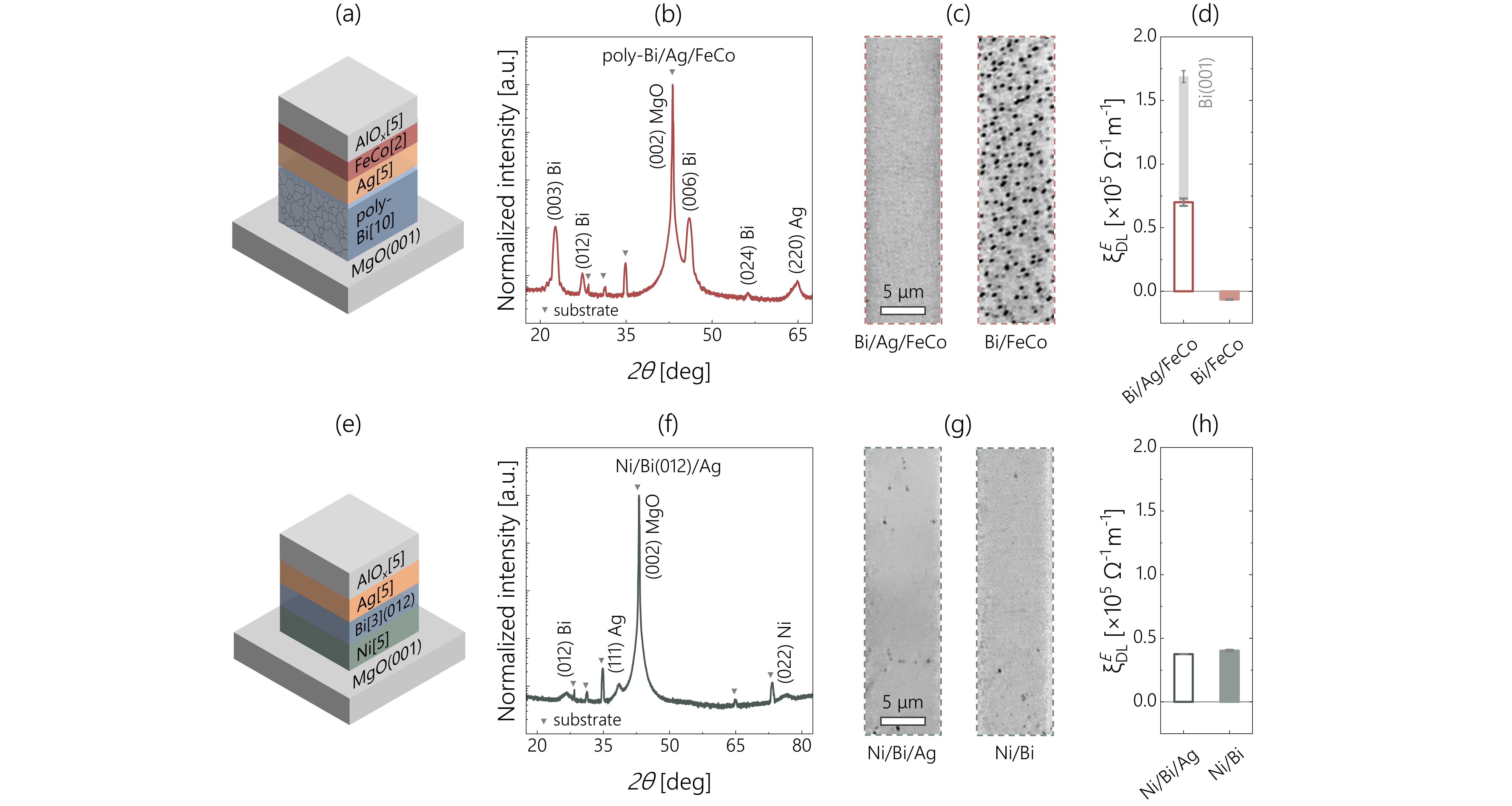}}
\caption{\label{fig:figure5} Structural characterization and SOT measurements of polycrystalline and (012)-textured Bi-based heterostructures grown on MgO(001).
(a, e) Schematic of the sample stacks, namely (a) poly-Bi[10]/Ag[5]/FeCo[2] and (e) Ni[5]/Bi[3](012)/Ag[5]. 
(b, f) XRD ${\theta/2\theta}$ scans of (b) poly-Bi[10]/Ag[5]/FeCo[2] and (f) Ni[5]/Bi[3](012)/Ag[5]. (c, g) Optical microscope images of a section of the Hall bar devices along the current line of (b) poly-Bi[10]/Ag[5]/FeCo[2] (left) and poly-Bi[10]/FeCo[2] (right); (f) Ni[5]/Bi[3](012)/Ag[5] (left) and Ni[5]/Bi[3](012) (right). (d, h) Damping-like SOT efficiencies ($\xi_\textrm{DL}^E$) with respect to the electric field ($E$) estimated using Eq.~\eqref{eq:xi_DL^E} for (d) poly-Bi[10]/Ag[5]/FeCo[2] and (h) Ni[5]/Bi[3](012)/Ag[5], with and without Ag. Error bars reflect the uncertainties of the weighted linear fits to the damping-like effective field ($B_\mathrm{DL}$) vs the average current density ($j$) with intercept forced to zero and of the saturation magnetic moment per unit area (${M_st_\mathrm{FM}}$). Positive SOT efficiencies indicate a positive SHC, consistent in sign with that of Pt. All thicknesses are expressed in [nm].}
\end{figure*}

To gain deeper insight into the origin of the SOT in the Bi/Ag bilayer, we further investigated the influence of Bi's structural properties on the magnitude and sign of the damping-like SOT. For this purpose, we fabricated additional heterostructures with distinct crystallographic configurations, namely, MgO(001)\slash\allowbreak Bi[10]\slash\allowbreak Ag[5]\slash\allowbreak FeCo[2]\slash\allowbreak AlO$_x$ and MgO(001)\slash\allowbreak Ni[5]\slash\allowbreak Bi[3]\slash\allowbreak Ag[5]\slash\allowbreak AlO$_x$, with and without Ag layer (see details in Sec.~\ref{sec:methods_growth}). Schematics of the sample stacks are provided in Figs.~\ref{fig:figure5}(a) and~\ref{fig:figure5}(e), respectively. 

\vspace{-0.3cm}
\subsubsection{Poly-Bi/Ag/FeCo heterostructure}
\vspace{-0.3cm}
In the former heterostructure\,$-$\,hereafter referred to as poly-Bi/Ag/FeCo\,$-$\,Bi grown on MgO(001) forms a polycrystalline film, composed of a mixture of (001) and (012)-oriented crystallites, as shown in the XRD ${\theta/2\theta}$ scan in Fig.~\ref{fig:figure5}(b). Although the Ag(110) texture is retained, the Laue oscillations associated with the (003) and (006) Bi reflexes are significantly damped compared to those observed in Bi(001)/Ag/FeCo, indicating increased structural roughness. This is further supported by the analysis of the (003) Bi peak: from its full width at half maximum, we estimate an out-of-plane crystallite size of approximately \SI{13.9}{nm} using the Scherrer equation (with a shape factor of 0.94). This value exceeds the nominal Bi thickness of \SI{10}{nm}, evidencing substantial surface roughness in the polycrystalline Bi film grown on MgO(001). Interestingly, similar to Bi(001)/FeCo, the poly-Bi/FeCo sample exhibits discontinuity in the Bi layer, with voids clearly visible in the optical microscopy image of Fig.~\ref{fig:figure5}(c). 

The SOT efficiencies of poly-Bi/Ag/FeCo and poly-Bi/FeCo are $\xi^E_\textrm{DL}$\,=\,$(0.70 \pm 0.03)$ and $(-0.064\pm 0.002)\times{10^5\,\SI{}{S/m}}$, respectively, as reported in Fig.~\ref{fig:figure5}(d). Notably, $\xi^E_\textrm{DL}$ in poly-Bi/Ag/FeCo is reduced by nearly 60\% compared to Bi(001)/Ag/FeCo, underscoring the strong sensitivity of the damping-like SOT to the structural and crystalline quality of Bi. This highlights the importance of epitaxial ordering and smooth interfaces in achieving efficient charge-to-spin conversion. Several mechanisms can contribute to the reduced SOT efficiency observed in the polycrystalline Bi film. Structural disorder, including grain boundaries and crystallographic defects, can induce variations in both the intrinsic SHC~\cite{Guo|JMMM|2022} and extrinsic skew-scattering contribution~\cite{Tatsuoka|PhysRevB|2026}. Moreover, the enhanced interdiffusion of different chemical species promoted by the higher density of grain boundaries can result in shifts of the electrochemical potential and modifications of the carrier population, to which the intrinsic SHC of Dirac-like semimetals such as Bi is known to be highly sensitive~\cite{Guo|JMMM|2022, Chi|SciAdv|2020}. Finally, poorer interface quality and local interfacial oxidation can reduce spin transparency for both bulk and interfacial spin currents~\cite{Manchon|RevModPhys|2019, Isshiki|PhysRevB|2020}, thus lowering the overall conversion efficiency. We note that the residual torque signal observed for poly-Bi/FeCo in Fig.~\ref{fig:figure5}(d) is comparable to that of Bi(001)/FeCo in Fig.~\ref{fig:figure4}(h). This supports the interpretation that, in the absence of an Ag spacer, the residual SOT primarily originates from self-torques within the FeCo layer.

\vspace{-0.3cm}
\subsubsection{Ni/Bi(012)/Ag heterostructure}
\vspace{-0.3cm}
The XRD pattern of the second heterostructure\,$-$\,hereafter referred to as Ni/Bi(012)/Ag\,$-$\,exhibits distinct reflections from the (022) crystal planes of Ni, the (012) planes of Bi, and the (111) planes of Ag, indicating that Bi is (012)-textured, as shown in Fig.~\ref{fig:figure5}(f). This crystallographic alignment between Bi(012) and Ag(111) is consistent with previous reports~\cite{Ast|PhysRevLett|2007}. In Ni/Bi(012)/Ag, we intentionally limited the thickness of Bi to \SI{3}{nm} to allow the spin angular momentum generated at the Bi/Ag interface to partially propagate into the underlying Ni layer, rather than being fully absorbed within Bi. This design choice is motivated by the findings of Ref.~\citenum{Fukumoto|PNAS|2023}, in which the spin-to-charge conversion in Ni/Bi(012) heterostructures saturates at a Bi thickness of approximately \SI{6}{nm}. Although a quantitative evaluation of the charge-to-spin conversion efficiency in Ni/Bi(012) lies beyond the scope of this study, we aim to assess whether the insertion of an Ag layer enhances the SOT signal of Bi, potentially via interfacial REE or \mbox{Bi--Ag} intermixing. The Bi films grown on Ni(011) appear continuous and uniform, as evidenced by the micrographs in Fig.~\ref{fig:figure5}(g). However, in contrast to the case where Ni is deposited on Bi(001), we observe no evidence of \mbox{Bi--Ni} alloying for this stacking configuration. Specifically, XRD measurements reveal no intermetallic diffraction peaks, and the Ni layer retains a robust magnetic response. The Oersted component of the Kerr rotation amplitude measured by MOKE is comparable to that of Bi(001)/Ag/Ni, under the same total applied current \{see Fig.~S4(b) within the Supplemental Material~\cite{SI}\}. Furthermore, saturation magnetization values extracted from SQUID magnetometry are $M_s$\,=\,\SI{3.33e5}{A/m} for Ni/Bi(012)/Ag and \SI{3.30E5}{A/m} for Ni/Bi(012), both closely matching that of Bi(001)/Ag/Ni of {\SI{3.54E5}{A/m}} \{see Fig.~S4(c) within the Supplemental Material~\cite{SI}\}. These findings indicate that the Ni layer in Ni/Bi(012) remains continuous and magnetically intact, in stark contrast to the significant interlayer diffusion and magnetic degradation observed when Ni is deposited on top of Bi(001). 

The SOT efficiencies of Ni/Bi(012)/Ag and Ni/Bi(012) are $\xi^E_\textrm{DL}$\,=\,$(0.38$\,$\pm$\,$0.01)$ and \mbox{$(0.40 \pm 0.01)$}\,$\times$\,$10^5\,\SI{}{S/m}$, respectively, as reported in Fig.~\ref{fig:figure5}(h). We note that the sign of $B_\text{DL}$ in these layers is opposite with respect to Bi(001)/Ag/Ni and Bi/Ni because of the reversed stacking order between the NM and the FM. The sign of $\xi^E_\textrm{DL}$, however, remains positive, consistent with our sign convention based on the SHC of Pt. Interestingly, Ni/Bi(012) exhibits a large SOT efficiency even without an Ag spacer between Bi and the FM and despite the Bi layer being only 3-nm-thick. Furthermore, depositing an Ag overlayer on top of Bi(012) has no measurable effect on the magnitude of the damping-like SOT. However, we cannot exclude the possibility of a spin-diffusion length in our Bi film much shorter than \SI{3}{\nm}, which would suppress any measurable contribution to the charge-to-spin conversion originating from the top Bi/Ag interface.

\section{Discussion}
\label{sec:discussion}
\vspace{-0.3cm}

We now compare the damping-like SOT efficiency across the various Bi-based heterostructures, correlating their structural and electronic properties with the corresponding torque magnitudes. Through this comparative analysis, we gain insights into the mechanisms underlying the charge-to-spin conversion in Bi and Bi/Ag bilayers. We further extract an effective SHA for Bi and benchmark our findings against previous literature reports on \mbox{spin--charge} interconversion in similar Bi and Bi/Ag system.

\vspace{-0.3cm}
\subsection{Correlation between SOTs and structural properties}
\vspace{-0.3cm}
The large and positive damping-like SOT observed in Bi(001)/Ag/FeCo, when compared to the small and negative torque signal measured in Bi(001)/FeCo and Ag/FeCo, at first glance appears to support the widely proposed scenario of a strong REE at the Bi/Ag interface. However, structural analysis based on HAADF-STEM and EDX reveals that the nominal Bi(001)/FeCo heterostructure effectively transforms into a Bi(001)/AlO$_x$/FeCo stack (see Sec.~\ref{sec:growth}). In this configuration, AlO$_x$ from the capping layer is unexpectedly detected at the Bi/FM interface, forming an insulating barrier that hinders spin transfer from Bi to FeCo. Moreover, the Bi layer exhibits significant oxygen incorporation and surface roughness, indicating that removal of the Ag spacer leads to deterioration of both the structural and chemical integrity of the Bi film. Although this observation does not rule out a contribution to the SOT from Rashba \mbox{spin--orbit} coupling at the Bi/Ag interface, it strongly underscores the critical role of the Ag spacer in preserving the Bi quality. Notably, this stabilizing effect is not replicated by other metallic spacers such as Cu or Al (see Sec.~\ref{subsec:spacers}). Further evidence for the detrimental impact of Bi oxidation comes from a reference sample in which Bi(001) was intentionally exposed to a controlled oxygen dose sufficient to degrade its surface states, following the experimental findings of Ref.~\citenum{Liu|PhysE|2012}. While a residual torque signal persists, suggesting a non-negligible bulk contribution to the charge-to-spin conversion, the overall magnitude of the damping-like SOT is substantially reduced (see Sec.~\ref{subsec:oxidation}). This observation is consistent with a previous report of time-dependent deterioration of the inverse SHE in an uncapped Bi layer owing to progressive oxidation~\cite{Liang|PhysRevB|2022}. Our study also reveals that the microscopic mechanisms driving changes in the compositional and microstructural properties of Bi films are strongly correlated with the mechanical stress and chemical environment imposed by adjoining layers. For instance, while no intermixing is observed when textured Bi(012) is deposited on Ni(011), reversing the stacking order\,$-$\,i.e., depositing Ni onto Bi(001)\,$-$\,induces significant interdiffusion and the formation of a Bi$_3$Ni intermetallic compound, which fundamentally alters the origin of the SOT (see Secs.~\ref{sec:structural_magnetic} and~\ref{sec:SOT_alternative}). These considerations also help elucidate the origin of the apparent anisotropy in the charge-to-spin conversion properties of our Bi films. In fact, the large damping-like SOT observed in Ni/Bi(012), as compared to the much smaller signals in Bi(001)/FM (see Secs.~\ref{sec:SOT} and ~\ref{sec:SOT_alternative}), can be largely explained by the extrinsic properties specific to each heterostructure, rather than to anisotropic transport effects intrinsic to the electronic structure of Bi. This interpretation is further supported by the observation that Bi(001) transforms into Bi(012) when capped by Ni. Although the anisotropy of the intrinsic SHC of Bi may partially contribute to the observed scaling of the SOT efficiency, assessing its role would require a comparison between samples with matched structural characteristics, conditions not satisfied by the current sample set.

These observations help rationalize inconsistencies previously reported in the literature. Most Bi/Ag bilayer studies are based on spin pumping, a characterization technique that relies on relatively thick ferromagnetic layers ($10$$-$$\SI{20}{nm}$) to ensure efficient and uniform ferromagnetic resonance excitation. Permalloy (Py, Ni$_{80}$Fe$_{20}$) is the most commonly used ferromagnet, and its Ni-rich composition makes it chemically prone to alloying with Bi. The small or negligible spin-to-charge conversion signals reported in several spin pumping studies~\cite{Sanchez|NatCommun|2013,Zhang|JApplPhys|2015,Nomura|ApplPhysLett|2015}, where Bi is directly interfaced with Py, may be explained by strong \mbox{Bi--Py} intermixing. Such intermixing could remain largely undetected since the bulk properties of the thick Py layer would be only weakly affected and thus retain their magnetic character. Moreover, capping Bi with thick FMs may obscure signatures of solid-state dewetting: instead of visible film breakup, dewetting may manifest only as an overall increase in surface roughness, unless depth-sensitive characterization techniques are employed.

Overall, our findings highlight the importance of exercising caution when interpreting \mbox{spin--charge} interconversion experiments in Bi-based heterostructures. Underlying solid-state dewetting and bulk/interfacial reactions are highly sensitive to the choice of ferromagnetic material, stacking order, interface chemistry, and growth conditions. Thickness is also a key factor governing the structural, electronic, and transport properties of Bi~\cite{Xiao|PhysRevLett|2012,Karbivskyy|ProgPhysMet|2021,Hirahara|PhysRevLett|2015}. A representative example is the large REE observed at the metallic surfaces of Bi nanocrystals, which is promoted by quantum confinement effects in ultrathin Bi films~\cite{Zucchetti|PhysRevB|2018}. Without careful structural and chemical characterization, such subtle yet crucial variations in sample configuration may remain undetected, leading to misleading comparisons among nominally similar stacks and contributing to the large variability of the reported \mbox{spin--charge} interconversion efficiencies across different studies.

\vspace{-0.3cm}
\subsection{Origin of SOTs}
\begin{figure*}[t!]
\centerline{\includegraphics[width=0.8\textwidth]{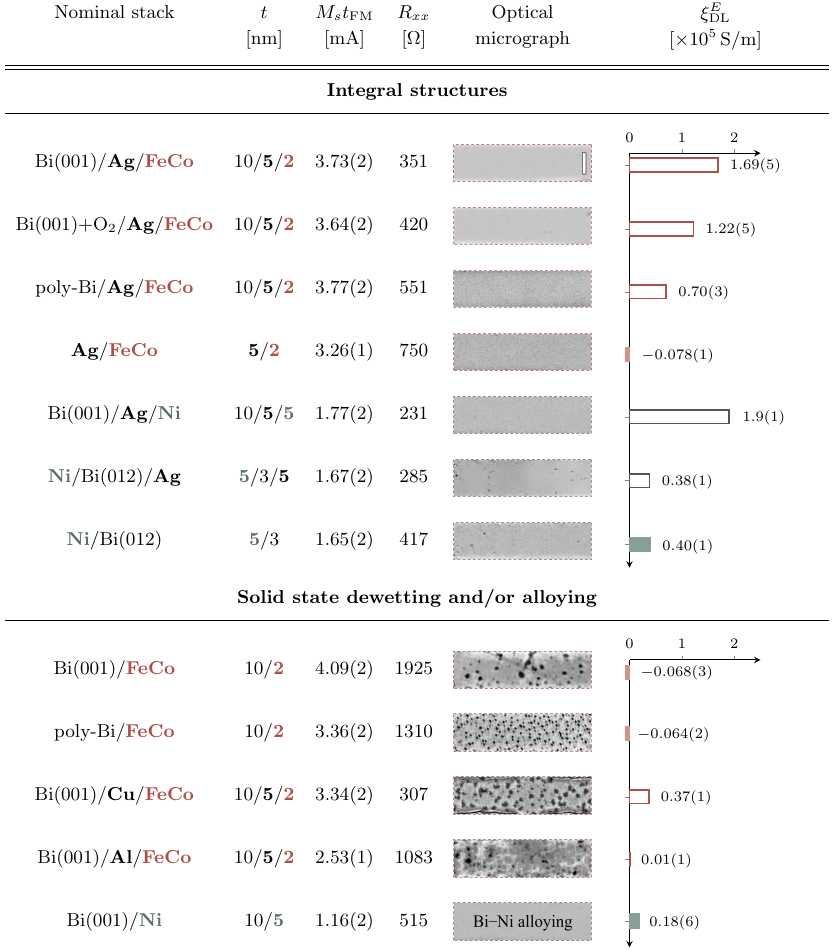}}
\caption{Summary of the structural, magnetic, electrical, and charge-to-spin conversion properties of the different samples. Shown in order are: nominal stack, layer thickness, saturation magnetization per unit area ($M_s\,t_\mathrm{FM}$), four-probe longitudinal resistance ($R^{L}$), optical micrograph, and horizontal bar plot of the effective SHC ($\xi^E_\textrm{DL}$). The sample stacks follow the bottom-to-top-layer convention. Red bars indicate samples with Fe$_{0.5}$Co$_{0.5}$, whereas green bars indicate samples with Ni as ferromagnetic layers. Positive $\xi^E_\textrm{DL}$ values correspond to a positive SHC of the nonmagnetic layer, consistent with that of Pt. All the stacks include an AlO$_x$[\SI{5}{nm}] cap. Numbers in parentheses denote the measurement uncertainty in the last digit.\label{fig:figure4}}
\label{fig:summary}
\end{figure*}

\vspace{-0.3cm}
ARPES measurements demonstrate that a long-range ordered BiAg$_2$ alloy does not form at the Bi(001)/Ag interface (see Sec.~\ref{sec:ARPES}). While ARPES alone cannot rule out the formation of a disordered alloy region promoting charge-to-spin conversion, HAADF-STEM imaging reveals a Bi/Ag interfacial region of approximately \SI{1}{nm} in Bi(001)/Ag/FeCo. This indicates that significant intermixing between Bi and Ag is unlikely in this sample (see Sec.~\ref{sec:growth}). Moreover, if a disordered interfacial \mbox{Bi--Ag} alloy was responsible for the large damping-like SOT in Bi(001)/Ag/FM, one would expect a significant enhancement of the SOT signal when Ag is deposited atop Bi in similar sample stacks. However, we observe no measurable change in damping-like SOT magnitude upon depositing Ag on Ni/Bi(012). Importantly, Ni/Bi(012) is the only sample in which the Bi layer retains its structural and chemical integrity without an Ag spacer, and it is also the only configuration in which Bi alone generates a robust damping-like SOT signal (see Sec.~\ref{sec:SOT_alternative}). Overall, the insertion of an Ag spacer enhances the SOT signal only when the integrity of the Bi layer would otherwise be compromised. To support this interpretation, we refer to the summary Fig.~\ref{fig:summary} which reveals a clear trend: continuous, chemically intact Bi layers consistently yield stronger charge-to-spin conversion, substantiating the role of Ag as an effective barrier against intermixing and a promoter of interface spin transparency. This conclusion parallels findings in (Bi$_{1-x}$Sb$_x$)$_2$Te$_3$, where intercalated normal metal spacers between the topological insulator and a permalloy layer are shown to reduce intermixing and spin memory loss, factors that can significantly alter the nature of SOT~\cite{Bonell|NanoLett|2020}. 
This interpretation is also consistent with the large SHAs of 0.3$-$0.4 recently measured for the Bi(001) semimetal and the topological insulators Bi$_{0.84}$Sb$_{0.16}$(001) and Bi$_{0.79}$Sb$_{0.21}$(001), in excellent agreement with tight binding calculations of the intrinsic SHC from bulk states~\cite{Ou|NanoLett|2025}.

\vspace{-0.3cm}
\subsection{Effective SHA and comparison with literature}
\vspace{-0.3cm}
Based on the above considerations, we extract the effective SHA of Bi under the assumption that the measured damping-like SOT in Bi(001)/Ag/FM arises from a combined contribution of bulk and surface effects within the Bi layer, and that Ag primarily acts as a structural stabilizer rather than a source of interfacial \mbox{spin--orbit} coupling. Within this framework, we compute $\xi_\textrm{DL}^{j_\textrm{Bi}}$\,=\,$\xi_\textrm{DL}^E/\sigma_\textrm{Bi}$, using the electrical conductivity of Bi $\sigma_\textrm{Bi}$\,=\,$\SI{1.57e5}{S/m}$ (corresponding to a resistivity of $\rho_\textrm{Bi}$\,=\,$\SI{635}{\micro\ohm \cm}$) estimated from a Bi(001) single layer capped with AlO$_x$. This yields $\xi_\textrm{DL}^{j_\textrm{Bi}}$\,=\,$1.2 \pm 0.2$, based on the average $\xi_\textrm{DL}^E$ obtained for Bi(001)/Ag/FeCo and Bi(001)/Ag/Ni. This value is in strong agreement with the average SHA of 1.3 reported previously for textured Bi(001)/Ag bilayers~\cite{Sanchez|NatCommun|2013}. It also closely matches the effective SHA of ${1.0\pm0.1}$ reported for Bi$_{0.9}$Sb$_{0.1}$(001)~\cite{Binda|AdvMater|2023}, with charge-to-spin conversion attributed to thermally excited holes from the bulk states rather than to topological surface states. The similarity in the calculated bulk band structures of Bi and Bi$_{0.9}$Sb$_{0.1}$~\cite{Sahin|PhysRevLett|2015}, as well as the comparable SOT efficiencies observed in these systems [both grown on BaF$_2$(111) under similar MBE conditions] point to a similar bulk-dominated origin of the SOT. The key distinction presumably lies in the stabilization strategy of the Bi-containing layer: while Ref.~\citenum{Binda|AdvMater|2023} achieved Bi stabilization through Sb alloying, in our case structural Bi integrity is preserved through the insertion of an Ag spacer. A more rigorous determination of the intrinsic SHA of Bi would require separating surface and bulk contributions to the measured charge-to-spin conversion, as well as accounting for possible self-induced torques in the Ni layer~\cite{Nasr|ApplPhyLett|2023}. We can therefore extract the effective SHA of Bi considering the SOT efficiency of Bi(001)+O$_2$/Ag/FeCo, which contains no Ni and for which oxygen exposure is expected to strongly suppress the Bi surface states. This yields $\xi_\textrm{DL}^{j_\textrm{Bi}}$\,=\,$0.78 \pm 0.03$, providing a lower bound for the bulk-dominated spin-current generation in Bi.

\section{Conclusions}
\label{sec:conclusion}
\vspace{-0.3cm}

We have systematically investigated the relationship between the structural properties and charge-to-spin conversion efficiency in Bi- and Bi/Ag-based magnetic heterostructures capped with AlO$_x$, including smooth epitaxial Bi(001) films as well as polycrystalline and (012)-textured layers with increased surface roughness. Our combined structural, spectroscopic, and MOKE-based SOT analyses reveal that the effective SHC $(\xi_\textrm{DL}^{E})$, expressed in units of [$\times$\,$10^5\,\SI{}{S/m}$] in the following, is primarily governed by the structural and chemical stability of the Bi layer, rather than by the REE at the Bi/Ag interface. 

When a 10-nm-thick Bi(001) film grown on BaF$_2$(111) is directly in contact with metallic FMs such as FeCo or Ni, we observe destabilization processes\,$-$\,including solid-state dewetting, roughening, oxidation, crystallographic reorientation, and interlayer mixing\,$-$\,that disrupt the intrinsic electronic structure of Bi and hinder effective angular momentum transfer to the FM. As a result, the nominal Bi(001)/FeCo stack exhibits small, negative, self-induced torques originating in the FeCo layer ($\xi_\textrm{DL}^{E}$\,=\,$-0.068 \pm 0.03$), effectively behaving as a partially oxidized Bi(001)/AlO$_x$/FeCo system. Likewise, the nominal Bi(001)/Ni stack develops spatially inhomogeneous and weak positive damping-like SOT ($\xi_\textrm{DL}^{E}$\,=\,$0.18 \pm 0.06$) because of its transformation into a Bi(012)/Bi$_3$Ni/Ni configuration. Introducing a thin Ag spacer between Bi and the FM successfully suppresses these degradation pathways, enabling large, positive, and spatially uniform damping-like SOT ($\xi_\textrm{DL}^{E}$\,=\,$1.69 \pm 0.05$ with FeCo and $1.9 \pm 0.1$ with Ni), consistent with theoretical prediction of bulk-mediated spin transport in semimetallic Bi. Extending the study to polycrystalline Bi grown on MgO(001) reveals similar degradation in the absence of an Ag spacer and an effective SHC reduction of nearly 60\% in poly-Bi/Ag/FeCo ($\xi_\textrm{DL}^{E}$\,=\,$0.70 \pm 0.03$) compared to Bi(001)/Ag/FeCo, reinforcing the strong sensitivity of the torque efficiency to the structural and crystalline quality of Bi.  

Further indications that the large charge-to-spin conversion signal in Bi(001)/Ag originates primarily from the preservation of Bi's bulk-like properties are provided by several complementary experimental observations. Unlike Ag, alternative metallic spacers such as Cu and Al fail to stabilize the Bi layer and yield significantly reduced or negligible damping-like SOT, underscoring the unique role of the Ag spacer as both efficient diffusion barrier and promoter of interface spin transparency. ARPES measurements demonstrate that Ag deposition on Bi(001) does not produce a long-range ordered \mbox{Bi--Ag} Rashba alloy but rather preserves the pristine band dispersion of Bi. Moreover, controlled oxygen exposure of the Bi(001) surface prior to Ag deposition reduces the effective SHC by approximately 28\%, yet leaves a significant residual signal ($\xi_\textrm{DL}^{E}$\,=\,$1.22 \pm 0.05$), consistent with the robustness of a bulk-dominated spin current generation mechanism. This yields a giant effective SHA for Bi of about 1. Finally, a 3-nm-thick Bi(012) film grown on a Ni(011) underlayer, despite lacking an Ag spacer, retains its structural continuity and shows no evidence of intermixing with Ni, in stark contrast to the nominal Bi(001)/Ni heterostructure. This inverted stack exhibits a sizable effective SHC ($\xi_\textrm{DL}^{E}$\,=\,$0.40 \pm 0.01$), which appears largely independent of the establishment of a Bi/Ag interface.

Taken together, these findings establish that the preservation of a structurally and chemically intact Bi layer is the key requirement for achieving large and reliable spin current generation and transmission. This provides a unified framework that helps reconcile long-standing discrepancies in the reported \mbox{spin--charge} interconversion efficiencies of Bi-based heterostructures. Future work should focus on refining epitaxial growth strategies, exploring alternative diffusion barriers with high spin transparency, and examining possible intrinsic anisotropies in Bi magnetotransport using structurally equivalent Bi films. Such efforts may pave the way toward robust Bi-based \mbox{spin--orbitronic} architectures that fully leverage the large intrinsic \mbox{spin--charge} interconversion capabilities of semimetallic Bi.

\clearpage

\begin{acknowledgments}
\vspace{-0.2cm}
We thank Dr. Ju-Young Yoon for his assistance with
scanning electron microscopy measurements and Drs. Shilei Ding, Min-Gu Kang, and Alexander Kossak for performing SQUID magnetometry measurements. We acknowledge financial support from the Swiss National Science Foundation (Grant No. 200021-236524). P.N.~acknowledges the support of the ETH Zurich Postdoctoral Fellowship Program (Grant No.~19-2 FEL-61). C.Z.~ acknowledges the support of IDEA League Fellowship 2024/2025. This work was partially funded through the project EUROFEL-ROADMAP ESFRI of the Italian MUR. We further acknowledge financial support from the European Commission through the Marie Sk\l{}odowska\mbox{-}Curie Actions H2020 RISE under the ULTIMATE-I project (Grant No. 101007825). We also thank the \textit{Laboratorio de Microscopías Avanzadas} and the \textit{Servicio General de Apoyo a la Investigación} at the Universidad de Zaragoza for providing access to microscopy facilities.
\end{acknowledgments}

\section*{Data availability}
\vspace{-0.3cm}
The data that support the findings of this study are openly available via the ETH Research Collection at https://doi.org/10.3929/ethz-c-000790287. 

\bibliography{main}

\end{document}